\documentclass[prb,aps,superscriptaddress,eqsecnum,twocolumn,amssymb,showpacs]{revtex4-1}

\usepackage{color}
\usepackage{graphicx}
\usepackage{longtable}
\usepackage{epsfig}
\usepackage{dcolumn}
\usepackage{bm}
\usepackage{amssymb}
\usepackage{amsmath}

\begin{document}

\title{Symmetry properties and spectra of the two-dimensional quantum compass model}

\author{Wojciech Brzezicki}
\affiliation{Marian Smoluchowski Institute of Physics, Jagellonian University,
Reymonta 4, 30-059 Krak\'ow, Poland }

\author{Andrzej M. Ole\'s }
\affiliation{Marian Smoluchowski Institute of Physics, Jagellonian University,
Reymonta 4, 30-059 Krak\'ow, Poland }
\affiliation{Max-Planck-Institut f\"ur Festk\"orperforschung,
Heisenbergstrasse 1, D-70569 Stuttgart, Germany }

\date{April 2, 2013}
\begin{abstract}
We use exact symmetry properties of the two-dimensional quantum compass
model to derive nonequivalent invariant subspaces in the energy spectra
of $L\times L$ clusters up to $L=6$. The symmetry allows one to reduce
the original $L\times L$ compass cluster to the $(L-1)\times (L-1)$ one
with modified interactions. This step is crucial and enables: (i) exact
diagonalization of the $6\times 6$ quantum compass cluster, and (ii)
finding the specific heat for clusters up to $L=6$, with two
characteristic energy scales. We investigate the properties of the
ground state and the first excited states and present extrapolation of
the excitation energy with increasing system size. Our analysis provides
physical insights into the nature of nematic order realized in the
quantum compass model at finite temperature. We suggest that the quantum
phase transition at the isotropic interaction point is second order with
some admixture of the discontinuous transition, as indicated by the
entropy, the overlap between two types of nematic order (on horizontal
and vertical bonds) and the existence of the critical exponent.
Extrapolation of the specific heat to the $L\to\infty$ limit suggests
the classical nature of the quantum compass model and high degeneracy
of the ground state with nematic order.
\end{abstract}


\maketitle

\section{Introduction}

Spin-orbital physics is a very exciting and challenging field of
research within the theory of strongly correlated electrons.
\cite{Nag00,Ole05,Kha05,Ole12}
Well known examples of Mott insulators with active orbital degrees of
freedom are two-dimensional (2D) and three-dimensional (3D) cuprates,
\cite{Fei97,*Fei98,Brz12,*Brz13} manganites,\cite{Fei99,*Fei05} and
vanadates.\cite{Kha01,*Hor08} These realistic models are rather
complicated and difficult to investigate due to spin-orbital
entanglement,\cite{Ole12} including the one on superexchange bonds.
\cite{Ole06,You12} A common feature of spin-orbital models is intrinsic
frustration of the orbital superexchange which follows from the
directional nature of orbital states and their interactions. The
orbital interactions are frequently considered alone, leading to
orbital ordered states,\cite{vdB99,vdB04,Ryn10} to valence bond crystal
or to orbital pinball liquid exotic quantum states.\cite{Ral12}

We shall concentrate below on a generic and the simplest model which
describes orbital-like superexchange, the so-called quantum compass
model (QCM),\cite{vdB13} introduced long ago by
Kugel and Khomskii.\cite{Kug82}
In this 2D model the coupling along a given bond is Ising-like, but
different spin components are active along different bond directions.
A frequently used convention is that interactions take the form
$J_{x}\sigma_{i}^{x}\sigma_{j}^{x}$ and
$J_{z}\sigma_{i}^{z}\sigma_{j}^{z}$ along $a$ and $b$ axis of the
square lattice.
The compass model is challenging already for classical interactions.
\cite{Nus04} Recent interest in this model is motivated by its
interdisciplinary character as it plays a role in the variety of
phenomena beyond the correlated oxides; is is also dual to recently
studied models of $p+ip$ superconducting arrays,\cite{Nus05} namely to
the Hamiltonian introduced by Xu and Moore,\cite{Xu04} and to the toric
code model in a transverse field.\cite{Vid09} Its 2D and 3D version
was studied in the general framework of unified approach to classical
and quantum dualities \cite{Cob10} and in the 2D case it was proved to
be self-dual.\cite{Xu04} The QCM was also suggested as an effective
description for Josephson arrays of protected qubits,\cite{Dou05} as
realized in recent experiment.\cite{Gla09} Finally, it could describe
polar molecules in optical lattices and systems of trapped ions.
\cite{Mil07}

First of all, the 2D QCM describes a quantum phase transition
between competing types of one-dimensional (1D) nematic orders, favored
either by $x$ or $z$ part of the Hamiltonian and accompanied by
discontinuous behavior of the nearest-neighbor (NN) spin correlations,
\cite{Kho03} when anisotropic interactions are varied through the
isotropic point $J_{x}=J_{z}$, as shown by high-order perturbation
theory,\cite{Mil05} rigorous mathematical approach,\cite{You10} mean
field (MF) theory on the Jordan-Wigner fermions,\cite{Che07} and
sophisticated infinite projected entangled-pair state (PEPS) algorithm.
\cite{Oru09} Thus, in the thermodynamic limit one of the involved
interactions is intrinsically frustrated because the energy of bonds
along one direction is minimized but the other is not. In fact, these
bonds which do not contribute to the actual spin order give no energy
gain and are totally ignored. Second, the quantum Monte-Carlo studies
of the isotropic QCM proved that the nematic order remains stable
at finite temperature up to $T_{c}=0.055J$ and the phase transition
to disordered phase stays in the Ising universality class.\cite{Wen08}
As shown by Dou\c{c}ot {\it et al.},\cite{Dou05} the eigenstates of the
QCM are twofold degenerate and the number of low-energy excitations
scales as linear size of the system. Further on, it was proved by exact
diagonalization of small systems that these excitations correspond to
the spin flips of whole rows or columns of the 2D lattice and survive
when a small admixture of the Heisenberg interactions is included into
the compass Hamiltonian.\cite{Tro10,*Tro12} The elaborated multiscale
entanglement-renormalization ansatz (MERA) calculations, together with
high-order spin wave expansion,\cite{Cin10} showed that the 2D QCM
undergoes a second order quantum phase transition when the interactions
are modified smoothly from the conflicting compass interactions towards
classical Ising model. It has also been shown\cite{Cin10} that the
isotropic QCM is not critical in the sense that the spin waves remain
gapful in the ground state, confirming that the order in the 2D QCM is
not of magnetic type.

For further discussion of the properties of the 2D QCM it is helpful
to recall the 1D case.
The 1D generalized variant of the compass model with $z$-th and $x$-th
spin component interactions that alternate on even/odd exchange bonds
is strongly frustrated, similar to the 2D QCM. The 1D QCM can be solved
exactly by an analytical method in two different ways.
\cite{wb07,wb09_act} We note that the 1D compass model is equivalent to
the 1D anisotropic XY model, solved exactly in the seventies.
\cite{Per75} An exact solution of the 1D compass model demonstrates that
certain NN spin correlation functions change discontinuously at the
point of a quantum phase transition (QPT) when both types of
interactions have the same strength, similarly to the 2D QCM. This
somewhat exotic behavior is due to the QPT occurring in this case at the
multicritical point in the parameter space.\cite{Eri09} The entanglement
measures, together with so called quantum discord in the ground state
characterizing the quantumness of the correlations, were analyzed
recently \cite{You08,You12} to find the location of quantum critical
points and to show that the correlations between two pseudospins on even
bonds are essentially classical in the 1D QCM. While small anisotropy of
interactions leads to particular short-range correlations dictated by
the stronger interaction, in both 1D and 2D compass model one finds a
QPT to a highly degenerate disordered ground state when the competing
interactions are balanced.

The purpose of this paper is to present the symmetry properties of the
2D compass model and their implications for the energy spectra. Exact
properties of the 2D QCM were introduced in Refs.
\onlinecite{wb10,wb10_icm,wb11_vi3}. Here we concentrate ourselves on
certain generic features and extensions which provide more insights into
the physical properties of the QCM. We present several results which
were not published until now --- they give a rather complete description
of the physical properties of the model. We apply the symmetry for
obtaining numerical results for the QCM on small square clusters,
including the $6\times 6$ cluster which becomes considerably easier
within the present approach than by a Lanczos exact diagonalization (ED)
in invariant subspaces of fixed $S^z$ which makes no use of the symmetry
described below. This symmetry is of importance here in spite of
remarkable progress in the ED studies performed recently on large
systems. For example, $S=\frac{1}{2}$ Heisenberg model was studied
recently on the kagome lattice with $N=42$ sites in the subspace with
$S^z=0$.\cite{Nak11}

The paper is organized as follows. In Sec. \ref{sec:sym} we focus first
on special symmetries of the planar QCM, giving the spin transformations
that bring the Hamiltonian into the block-diagonal (or reduced) form and
confirm its self-duality (Sec. \ref{sub:block}). Next, in Sec.
\ref{sub:sub}, we derive the equivalence relations between these diagonal
blocks (or invariant subspaces) following from the translational
invariance of the original QCM Hamiltonian and show the multiplet
structure of the invariant subspaces for $4\times 4$, $5\times 5$ and
$6\times 6$ lattices in Sec. \ref{sub:multi} and in the Appendix. The
study of symmetries culminates in unveiling the hidden symmetry of the
ground state of the QCM, see Sec. \ref{sub:hidden}, and its consequences
for the four-point correlation functions using another spin
transformation. Next, in Sec. \ref{sec:num}, we present the results of
ED techniques applied to the QCM for lattices of
the sizes up to $6\times6$. Due to the complexity of the many-body
problem which includes time-consuming implementation of symmetry
properties of the 2D QCM, this can be regarded as the state-of-the-art
implementation of ED, see Sec. \ref{sub:kpm}. The results include ground
state properties of the QCM such as: spin correlation functions and
covariances of the local and nonlocal type in Sec. \ref{sub:gs}. In Sec.
\ref{sub:ele} we present the evolution of energy levels as a functions
of anisotropy, and entanglement entropy of a row in the lattice. We
study as well the density of states for the $6\times 6$ cluster and heat
capacities of the systems of different size at the isotropic point, see
Sec. \ref{sub:cv}. The paper is summarized briefly in Sec. \ref{sec:summa}.

\section{Symmetry properties of the two-dimensional compass model}
\label{sec:sym}

\subsection{Block-diagonal Hamiltonian}
\label{sub:block}

We consider the anisotropic ferromagnetic QCM for pseudospins $\frac12$
on a finite $L\times L$ square lattice with periodic boundary
conditions (PBCs):
\begin{eqnarray}
{\cal H}(\alpha)&=&-J\sum_{i,j=1}^{L}\left\{
(1-\alpha)X_{i,j}X_{i+1,j}+\alpha Z_{i,j}Z_{i,j+1}\right\}\nonumber\\
&=& -(1-\alpha)H^x-\alpha H^z\,,
\label{ham}
\end{eqnarray}
where $\{X_{i,j},Z_{i,j}\}$ stand for Pauli matrices at site $(i,j)$ of
a 2D square lattice,
i.e., $X_{i,j}\equiv\sigma_{i,j}^x$ and $Z_{i,j}\equiv\sigma_{i,j}^z$
components, interacting on vertical and horizontal bonds by $H^x$ and
$H^z$, respectively. The coupling constant $J$ is positive and the sign
factor $-1$ is introduced to provide comparable ground state properties
for odd and even systems. In this section we set $J=1$. The parameter
$\alpha\in[0,1]$ changes the anisotropy between horizontal ($H^x$) and
vertical ($H^z$) interactions;
the isotropic model is found at $\alpha=\frac12$. In case of $L$ being
even, this model is equivalent to the antiferromagnetic QCM.

One can easily construct a set of $2L$ operators which commute
with the Hamiltonian but anti-commute with one another:\cite{Dou05}
\begin{equation}
P_{i}\equiv\prod_{j=1}^{L}\, X_{i,j}, \hskip .7cm
Q_{j}\equiv\prod_{i=1}^{L}\, Z_{i,j}.
\end{equation}
Below we will use as symmetry operations all
\begin{equation}
R_{i}\equiv P_{i}P_{i+1}
\end{equation}
and $Q_{j}$ to reduce the Hilbert space; this approach led to the
exact solution of the compass ladder.\cite{wb09} The QCM Eq. (1)
can be written in a common eigenbasis of $\{R_{i},Q_{j}\}$ operators
using spin transformations of the form:
\begin{eqnarray}
X_{i,j} & = & \prod_{p=i}^{L}\tilde{X}_{p,j}\,,\hskip1cm
\tilde{X}_{i,j}=X'_{i,j-1}X'_{i,j}\,,\label{transx}\\
Z_{i,j} & = & \tilde{Z}_{i-1,j}\tilde{Z}_{i,j}\,,\hskip.7cm
\tilde{Z}_{i,j}=\prod_{q=j}^{L}Z'_{i,q}\,,\label{transz}
\end{eqnarray}
where $\tilde{Z}_{0,j}\equiv1$ and $X'_{i,0}\equiv1$. After writing
the Hamiltonian ${\cal H}(\alpha)$ of Eq. (\ref{ham}) in terms of
primed pseudospin operators one finds that the transformed Hamiltonian,
\begin{equation}
{\cal H}'(\alpha)=-(1-\alpha)H'_{x}-\alpha H'_{z},
\end{equation}
contains no $\tilde{X}_{L,j}$ and no $Z'_{i,L}$ operators so the
corresponding $\tilde{Z}_{L,j}$ and $X'_{i,L}$ can be replaced by
their eigenvalues $q_{j}$ and $r_{i}$, respectively.

The Hamiltonian ${\cal H}'(\alpha)$ is dual to the QCM
${\cal H}(\alpha)$ Eq. (\ref{ham}) in the thermodynamic limit;
we give here an explicit form of its transformed $x$-part:
\begin{eqnarray}
H'_{x} & = & \sum_{i=1}^{L-1}\left\{ \sum_{j=1}^{L-2}X'_{i,j}X'_{i,j+1}
+X'_{i,1}+r_{i}X'_{i,L-1}\right\} \nonumber \\
 & + & P'_{1}+\sum_{j=1}^{L-2}P'_{j}P'_{j+1}+rP'_{L-1},
 \label{Hiks}
\end{eqnarray}
and the similar form for the $z$-part:
\begin{eqnarray}
H'_{z} & = & \sum_{j=1}^{L-1}\left\{ \sum_{i=1}^{L-2}Z'_{i,j}Z'_{i+1,j}
+Z'_{1,j}+s_{j}Z'_{L-1,j}\right\} \nonumber \\
 & + & Q'_{1}+\sum_{j=1}^{L-2}Q'_{i}Q'_{i+1}+sQ'_{L-1},
\label{Hzet}
\end{eqnarray}
where $s_{j}=q_{j}q_{j+1}$, $s=\prod_{j=1}^{L-1}s_{j}$ and
$r=\prod_{i=1}^{L-1}r_{i}$,
and new nonlocal operators,
\begin{equation}
P'_{j}=\prod_{p=1}^{L-1}\, X'_{p,j}, \hskip .7cm
Q'_{i}=\prod_{q=1}^{L-1}\, Z'_{i,q},
\end{equation}
originate from the PBCs. As we can see, the $z$-th part $H'_{z}$
(\ref{Hzet}) follows from $H'_{x}$ (\ref{Hiks}) by the lattice
transposition, replacing $X'_{i,j}\rightarrow Z'_{i,j}$ and
$r_{i}\rightarrow s_{j}=q_{j}q_{j+1}$.
Ising variables $r_{i}$ and $s_{j}$ are the eigenvalues of the symmetry
operators $R_{i}\equiv P_{i}P_{i+1}$ and $S_{j}=Q_{j}Q_{j+1}$.

Instead of the initial $L\times L$ lattice of quantum spins, one
finds here $(L-1)\times(L-1)$ internal quantum spins with $2(L-1)$
classical boundary spins, which gives $L^{2}-1$ degrees of freedom.
The missing spin is related to the $Z_{2}$ symmetry of the QCM and
makes every energy level at least doubly degenerate. Although the form
of Eqs. (\ref{Hiks}) and (\ref{Hzet}) is complex, the size of the
Hilbert space is reduced in a dramatic way\cite{wb10_vi2,wb10_icm} by
a factor $2^{2L-1}$ which makes it possible to perform easily exact
(Lanczos) diagonalization of 2D $L\times L$ clusters up to $L=6$.

\begin{figure}[t!]
\begin{center}\includegraphics[width=8cm]{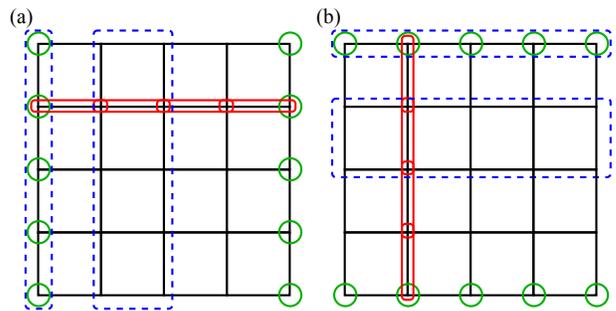}
\end{center}
\caption{Panel (a): Schematic view of the $x$-th part of the effective
compass Hamiltonian $H'_{x}$ (\ref{Hiks}): open (green) circles are
$X'_{i,j}$ spin operators acting along first and last column, dashed
(blue) frames indicate nonlocal $P'_{j}$ spin operator products along
columns and solid (red) frames are NN compass bonds $X'_{i,j}X'_{i,j+1}$.
Panel (b): Schematic view of $H'_{z}$ (\ref{Hzet}): open (green) circles
are $Z'_{i,j}$ spin operators acting along first and last line, dashed
(blue) frames symbolize nonlocal $Q'_{i}$ spin operator products along
lines and solid (red) frames are NN compass bonds $Z'_{i,j}Z'_{i+1,j}$.}
\label{fig:cmpefhamsprim}
\end{figure}

\subsection{Equivalent subspaces}
\label{sub:sub}

The spin transformations defined by Eqs. (\ref{transx}) and
(\ref{transz}) bring the QCM Hamiltonian (\ref{ham}) into the
block-diagonal form of Eqs. (\ref{Hiks}) and (\ref{Hzet}) with invariant
subspaces labeled by the pairs of vectors $(\vec{r},\vec{s})$, with
$\vec{r}=(r_{1},r_{2},\dots,r_{L-1})$ and
$\vec{s}=(s_{1},s_{2},\dots,s_{L-1})$. The original QCM of Eq.
(\ref{ham}) is invariant under the transformation $X\leftrightarrow Z$,
if one also transforms the interactions,
$\alpha\leftrightarrow(1-\alpha)$. This sets a relation between
different invariant subspaces $(\vec{r},\vec{s})$, i.e., after
transforming $\alpha\leftrightarrow(1-\alpha)$ the QCM Hamiltonian in
subspaces $(\vec{r},\vec{s})$ and $(\vec{s},\vec{r})$ has the same
energy spectrum. In general, we may say that the two subspaces are
equivalent if the QCM has in them the same energy spectrum. This
relation becomes especially simple for $\alpha=\frac{1}{2}$ when
for all $r_{i}$'s and $s_{i}$'s subspaces $(\vec{r},\vec{s})$ and
$(\vec{s},\vec{r})$ are equivalent.

Now we will explore another important symmetry of the 2D compass model
reducing the number of nonequivalent subspaces --- the translational
symmetry. We note from Eqs. (\ref{Hiks}) and (\ref{Hzet}) that the
reduced Hamiltonians are not translationally invariant for any choice of
$(\vec{r},\vec{s})$ even though the original Hamiltonian is. This means
that translational symmetry must impose some equivalence conditions
among subspace labels $(\vec{r},\vec{s})$. To derive them, let us focus
on translation along the rows of the lattice by one lattice constant.
Such translation does not affect the $P_{i}$ symmetry operators, because
they consist of spin operators multiplied along the rows, but changes
$Q_{j}$ into $Q_{j+1}$ for all $j<L$ and $Q_{L}\rightarrow Q_{1}$. This
implies that two subspaces $(\vec{r},q_{1},q_{2},\dots,q_{L})$ and
$(\vec{r},q_{L},q_{1},q_{2},\dots,q_{L-1})$ are equivalent for all
values of $\vec{r}$ and $\vec{q}$.

Now this result must be translated
into the language of $(\vec{r},\vec{s})$ labels, with
$s_{j}=q_{j}q_{j+1}$ for all $j<L$. This is two-to-one mapping because
for any $\vec{s}$ one has two $\vec{q}$'s such that:
\begin{eqnarray}
\vec{q}_{+}&=&(1,s_{1},s_{1}s_{2},\dots,s_{1}s_{2}\dots s_{L-1}),\nonumber\\
\vec{q}_{-}&=&-\vec{q}_{+}\,.
\label{qq}
\end{eqnarray}
The two values $\{q_+,q_-\}$ differ by global inversion. This sets
additional equivalence condition for subspace labels $(\vec{r},\vec{s})$:
two subspaces $(\vec{r},\vec{u})$ and $(\vec{r},\vec{v})$ are equivalent
if two strings $(1,u_{1},u_{1}u_{2},\dots,u_{1}u_{2}\dots u_{L-1})$
and $(1,v_{1},v_{1}v_{2},\dots,v_{1}v_{2}\dots v_{L-1})$ are related
by translations or by a global inversion. For convenience, let us call
this property of the two vectors a translation inversion (TI) relation.
Lattice translations along the columns set the same equivalence
condition for $\vec{r}$ labels. Thus full equivalence conditions for
subspace labels of the QCM are:
\begin{itemize}
\item For $\alpha=\frac{1}{2}$ two subspaces $(\vec{r},\vec{s})$ and
$(\vec{u},\vec{v})$
are equivalent if $\vec{r}$ is TI-related with $\vec{u}$ and $\vec{s}$
with $\vec{v}$ or if $\vec{r}$ is TI-related with $\vec{v}$ and
$\vec{s}$ with $\vec{u}$.
\item For $\alpha\not=\frac{1}{2}$ two subspaces $(\vec{r},\vec{s})$ and
$(\vec{u},\vec{v})$ are equivalent if $\vec{r}$ is TI-related with
$\vec{u}$ and $\vec{s}$ with $\vec{v}$.
\end{itemize}
We have verified that no other equivalence conditions exist between
the subspaces by numerical Lanczos diagonalizations for lattices of
sizes up to $6\times6$, so we can change all \textit{if\/} statements
above into \textit{if and only if\/} ones.

\section{Consequences of symmetry}
\label{sec:con}

\subsection{Multiplets of equivalent subspaces: examples}
\label{sub:multi}

For the finite square clusters of the sizes $4\times 4$, $5\times 5$
and $6\times 6$ we used the reduced form of the compass Hamiltonian to
reduce the dimensionality of the Hilbert space and apply exact
diagonalization techniques to get the ground state and thermodynamic
properties of the QCM. For this purpose we needed to create a list of
inequivalent subspaces for $L=4,5,6$ to save time and computational
effort. According to the previous discussion let's denote all
inequivalent $\vec{r}$ configurations for our systems.
For $L=4$ these fall into four TI-equivalence classes,
\[
\begin{array}{cccc}
\{\left[-+++\right], & \left[--++\right], & \left[-+-+\right], &
\left[----\right]\}\end{array},
\]
where the sign labels $q_+$ and $q_-$ in Eqs. (\ref{qq}), respectively.
the number of different $\vec{q}$ labels that can be constructed out of
each class is equal to the cardinality of this class divided by two.
For the $4\times 4$ system these numbers are $\{4,2,1,1\}$. For
$\vec{p}$ labels we have exactly the same set of classes so the subspace
structure can be characterized by the following diagrams,
\begin{equation}
\begin{array}{cc}
\begin{array}{cccc}
16 & 8 & 4 & 4\\
8 & 4 & 2 & 2\\
4 & 2 & 1 & 1\\
4 & 2 & 1 & 1
\end{array}\,\,, & \hskip 2cm
\begin{array}{cccc}
16 & 16 & 8 & 8\\
 & 4 & 4 & 4\\
 &  & 1 & 2\\
 &  &  & 1
\end{array}\,\,,\end{array}
\label{eq:diag16}
\end{equation}
where each number symbolizes an equivalence class of subspaces in
anisotropic (left) and isotropic (right) cases and is equal to the
number of subspaces in each class divided by two. As we see, the right
diagram can be obtained from the left one by leaving diagonal numbers
untouched, removing subdiagonal numbers and doubling the upper ones.

For $5\times 5$ we have again four TI--equivalence classes,
\[
\{\left[-++++\right],\left[--+++\right],\left[-+-++\right],\left[-----\right]\},
\]
with half-cardinalities $\{5,5,5,1\}$. This leads to the following
diagrams,
\begin{equation}
\begin{array}{cc}
\begin{array}{cccc}
25 & 25 & 25 & 5\\
25 & 25 & 25 & 5\\
25 & 25 & 25 & 5\\
5 & 5 & 5 & 1
\end{array}\,\,, & \hskip 2cm
\begin{array}{cccc}
25 & 50 & 50 & 10\\
 & 25 & 50 & 10\\
 &  & 25 & 10\\
 &  &  & 1
\end{array}\,\,.\end{array}
\label{eq:diag25}
\end{equation}

Finally, for the largest system considered here of $L=6$,
the TI--equivalence classes read,
\begin{equation}
\begin{array}{cccc}
\{\left[-+++++\right], & \left[--++++\right], & \left[-+-+++\right],\\
\left[--++-+\right], & \left[---+++\right], & \left[-++-++\right],\\
\left[+-+-+-\right], & \left[------\right]\},
\end{array}\label{eq:pm}
\end{equation}
with half-cardinalities $\{6,6,6,6,3,3,1,1\}$, yielding the anisotropic
diagram of the form,
\begin{equation}
\begin{array}{cccccccc}
36 & 36 & 36 & 36 & 18 & 18 & 6 & 6\\
36 & 36 & 36 & 36 & 18 & 18 & 6 & 6\\
36 & 36 & 36 & 36 & 18 & 18 & 6 & 6\\
36 & 36 & 36 & 36 & 18 & 18 & 6 & 6\\
18 & 18 & 18 & 18 & 9 & 9 & 3 & 3\\
18 & 18 & 18 & 18 & 9 & 9 & 3 & 3\\
6 & 6 & 6 & 6 & 3 & 3 & 1 & 1\\
6 & 6 & 6 & 6 & 3 & 3 & 1 & 1
\end{array}\,.
\label{eq:diag36}
\end{equation}
The isotropic diagram can be obtained using the known procedure. These
examples show that the number of inequivalent subspaces $N$ stays the
same for the systems of sizes $L=2l$ and $L=2l+1$ (with $l=1,2,3,\dots$)
and is directly related to the number $n$ of TI--equivalence classes
of the binary string of the length $L$. We have:
\begin{equation}
N=\begin{cases}
n^{2} & for\qquad\;\alpha\neq\frac{1}{2}, \\
\frac{1}{2}n(n+1) & for\qquad\;\alpha=\frac{1}{2}.
\end{cases}
\end{equation}

The most numerous TI-equivalence class for the $L\times L$ system
consists of the binary strings which transform into themselves after
$L$ translations, so carrying highest number of possible pseudomomenta.
This implies that largest subspace equivalence class contains $2L^{2}$
subspaces in anisotropic and $4L^{2}$ subspaces in isotropic case.
Knowing that the total number of subspaces is $2\times 2^{2(L-1)}$ one
can estimate that $N>2^{2(L-1)} L^{-2}$ for $\alpha\not=\frac{1}{2}$,
and $N>2^{2L-3} L^{-2}$ for $\alpha=\frac{1}{2}$.

\subsection{Hidden order}
\label{sub:hidden}

\begin{figure}[t!]
\begin{center}\includegraphics[width=8cm]{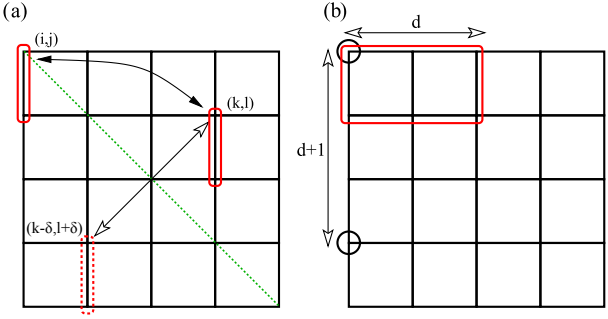}
\end{center}
\caption{Example of application of the proved identities in two cases:
(a) --- Eq. (\ref{mult}) long range correlation function
$\langle X_{i,j}X_{i+d+1,j}\rangle$ along the column (circles) is equal
to the $2d$--point $\langle XX\dots X\rangle$ correlation function
along the row (solid (red) frame of length $d$);
(b) --- Eq. (\ref{fin}) for two chosen dimers at $(i,j)$ and $(k,l)$
(solid frames), correlations between them are the same as between
dimers at $(i,j)$ and $(k-\delta,l+\delta)$ (dashed frame). Green
dashed line marks the plane of the mirror reflection transforming
site $(k,l)$ into $(k-\delta,l+\delta)$.}
\label{fig:appl}
\end{figure}

Due to the symmetries of the QCM Eq. (\ref{ham}) only
$\langle Z_{i,j}Z_{i,j+d}\rangle$ and $\langle X_{i,j}X_{i+d,j}\rangle$
two-point spin correlations are finite ($d\geq1$). This suggests that
the entire spin order concerns \textit{pairs of spins\/} from one row
(column) which could be characterized by four--point correlation
functions of the dimer-dimer type.
Such correlations are presented in form of the $2d$ point
$\langle XXXX\rangle$ correlation function in Fig. \ref{fig:appl}(a) and
by dimer-dimer correlations in Fig. \ref{fig:appl}(b).
Indeed, examining such quantities
for finite QCM clusters via Lanczos diagonalization we observed certain
surprising symmetry: for any $\alpha$ two dimer-dimer
$\langle (XX)_1(XX)_2\rangle$ correlators are equal
under the quasi-reflection along the local diagonal
as shown in Fig. \ref{fig:appl}(b).
This property turns out to be a special case of a more general relation
between correlation functions of the QCM which we prove below.

We will prove that in the ground state of the QCM for any two sites
$(i,j)$ and $(k,l)$ and for any $\alpha\in(0,1)$:
\begin{eqnarray}
\langle X_{i,j}X_{i+1,j}X_{k,l}X_{k+1,l}\rangle\equiv\quad\quad\quad
\nonumber \\
\langle X_{i,j}X_{i+1,j}X_{l-\delta,k+\delta}X_{l-\delta+1,k+\delta}\rangle,
\label{thes}
\end{eqnarray}
where $\delta=j-i$. To prove Eq. (\ref{thes}) let us transform again
the effective Hamiltonian (\ref{Hiks}) in the ground state subspace
($r_{i}\equiv s_{i}\equiv1$) introducing new spin operators,
\begin{equation}
Z'_{i,j}=\tilde{Z}_{i,j}\tilde{Z}_{i,j+1},\hskip 1cm
X'_{i,j}=\prod_{r=1}^{j}\tilde{X}_{i,r},
\label{trans2}
\end{equation}
with $i,j=1,\dots,L-1$ and $\tilde{Z}_{i,L}\equiv1$. This yields
\begin{equation}
\tilde{H}_{x}=\sum_{i,j=1}^{L-1}\!\tilde{X}_{i,j}
+\prod_{i,j=1}^{L-1}\!\tilde{X}_{i,j}+\sum_{i=1}^{L-1}
\prod_{j=1}^{L-1}\!\tilde{X}_{i,j}+\sum_{i=1}^{L-1}
\prod_{j=1}^{L-1}\!\tilde{X}_{j,i},\label{Hiks2}
\end{equation}
and
\begin{eqnarray}
\tilde{H}_{z} & = & \sum_{a}\left\{ \sum_{b}\tilde{Z}_{a,b}
+\sum_{i=1}^{L-2}\left(\tilde{Z}_{a,i}\tilde{Z}_{a,i+1}
+\tilde{Z}_{i,a}\tilde{Z}_{i+1,a}\right)\right\} \nonumber \\
 & + & \sum_{i=1}^{L-2}\sum_{j=1}^{L-2}\tilde{Z}_{i,j}\tilde{Z}_{i,j+1}
\tilde{Z}_{i+1,j}\tilde{Z}_{i+1,j+1},\label{Hzet2}
\end{eqnarray}
where $a=1,L-1$ and $b=1,L-1$. Due to the spin transformations Eqs.
(\ref{transx}), (\ref{transz}), and (\ref{trans2}),
$\tilde{X}_{i,j}$ operators are related to the original bond operators
by $X_{i,j}X_{i+1,j}=\tilde{X}_{i,j}$, which implies that
\begin{equation}
\langle X_{i,j}X_{i+1,j}X_{k,l}X_{k+1,l}\rangle=
\langle\tilde{X}_{i,j}\tilde{X}_{k,l}\rangle.
\end{equation}

\begin{figure}[t!]
\begin{center}
\includegraphics[width=8cm]{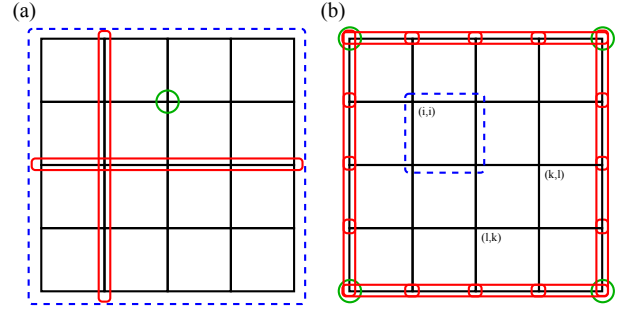}
\end{center}
\caption{Panel (a): Schematic view of the $x$-th part of the reduced
ground state subspace Hamiltonian $\tilde{H}_{x}$ (\ref{Hiks2}): empty
(green) circles are $\tilde{X}_{i,j}$ spin operators acting on every
site, dashed (blue) frame symbolize nonlocal product of all
$\tilde{X}_{i,j}$ operators products and solid (red) frames are products
of $\tilde{X}_{i,j}$ along all lines and columns. Panel (b): Schematic
view of $\tilde{H}_{z}$ (\ref{Hzet2}): empty (green) circles in the
corners stand for $\tilde{Z}_{i,j}$ spin operators related to the site
$(i,j)$, solid (red) frames are $\tilde{Z}\tilde{Z}$ operator products
acting on the boundaries of the lattice, and dashed (blue) square stands
for one of the plaquette $\tilde{Z}\tilde{Z}\tilde{Z}\tilde{Z}$ spin
operators. The exemplary three sites in the identity (\ref{fin}) are:
$(i,i)$, $(k,l)$ and $(l,k)$.}
\label{fig:cmpefhamstilde}
\end{figure}

Because of the PBC, all original $X_{i,j}$ spins are equivalent, so we
choose $i=j$. The $x$-part (\ref{Hiks2}) of the Hamiltonian is
completely isotropic. Note that the $z$-part (\ref{Hzet2}) would also be
isotropic without the boundary terms (see Fig. \ref{fig:cmpefhamstilde});
the effective Hamiltonian in the ground subspace has the symmetry
of a square. Knowing that in the ground state we have only $Z_{2}$
degeneracy, one finds
\begin{equation}
\langle\tilde{X}_{i,i}\tilde{X}_{k,l}\rangle\equiv\langle
\tilde{X}_{i,i}\tilde{X}_{l,k}\rangle,\label{fin}
\end{equation}
for any $i$ and $(k,l)$. This proves the identity (\ref{thes})
for $\delta=0$; $\delta\neq0$ case follows from lattice translations
along rows.

\begin{figure}[t!]
\begin{center}
\includegraphics[width=7.5cm]{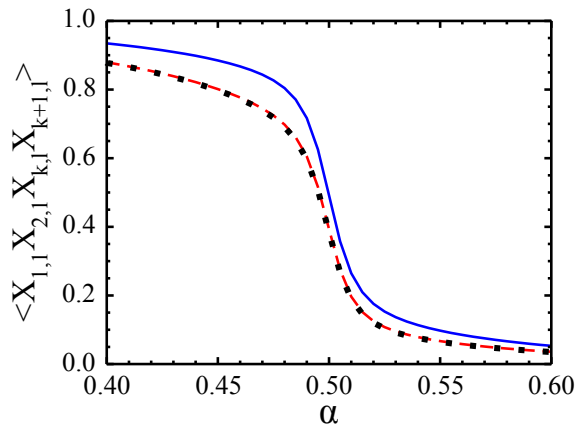}
\end{center}
\caption{Dimer-dimer correlations
$\langle X_{1,1}X_{2,1}X_{k,l}X_{k+1,l}\rangle$ for $L=6$ and
$0.4<\alpha<0.6$: $(k,l)=(1,2)$, $(1,3)$ and $(1,4)$ are shown by
solid, dashed and dotted line, respectively. }
\label{fig:didi}
\end{figure}

The nontrivial consequences of Eq. (\ref{fin}) are:
(i) \textit{hidden dimer order\/} in the ground state of the QCM ---
dimer correlation functions in two \textit{a priori\/} nonequivalent
directions in the QCM are identical and robust for $X$-components for
$\alpha<\frac12$ (Fig. \ref{fig:didi}) and for $Z$-components for
$\alpha>\frac{1}{2}$ (not shown), and
(ii) long range two-site $\langle X_{i,j}X_{i+d+1,j}\rangle$
correlations along the columns which are equal to the multi-site
$\langle XX\dots X\rangle$ correlations involving two neighboring rows,
see Fig. \ref{fig:appl}(a). The latter comes from symmetry
properties of the transformed Hamiltonian Eqs. (\ref{Hiks2}) and
(\ref{Hzet2}) applied to the multi-site correlations:
\begin{equation}
\langle\tilde{X}_{i,i}\tilde{X}_{i,i+1}\dots\tilde{X}_{i,i+d}\rangle=
\langle\tilde{X}_{i,i}\tilde{X}_{i+1,i}\dots\tilde{X}_{i+d,i}\rangle.
\label{mult}
\end{equation}

\section{Numerical studies on finite square clusters}
\label{sec:num}

\subsection{Exact diagonalization methods}
\label{sub:kpm}

Although there is no exact solution for the 2D QCM Eq. (\ref{ham}), the
latest Monte Carlo data \cite{Wen08} prove that the model exhibits a
phase transition at finite temperature both in quantum and classical
version, with symmetry breaking between $x$ and $z$ part of the QCM.
In this section we suggests a scenario for a phase transition with
increasing cluster size by the behavior of spin-spin correlation
functions and von Neumann entropy of a single column in the ground
state obtained via Lanczos algorithm. We also present the specific heat
calculated using Kernel Polynomial Method (KPM).\cite{Feh08}

Ground state energies and energy gap of the 2D QCM has already been
calculated for different values of $\alpha$
and for square $L\times L$ clusters with $L\in[2,5]$ using ED and for
higher $L$ using Green's function Monte Carlo method.\cite{Mil05}
Our approach is based on Lanczos algorithm and KPM \cite{Feh08} which
lets us calculate the densities of states and the partition functions
for square lattices of the sizes up to $L=6$. We start by applying
Lanczos algorithm to determine spectrum width which is needed for
KPM calculations. The resulting few lowest energies that we get from
the Lanczos recursion can be compared with the density of states to
check whether the KPM results are correct.

One should be aware that the
spectra of odd systems are qualitatively different from those of even
ones. For the even systems operator $S$ defined as
\begin{equation}
S=\prod_{i,j=1}^{L}\frac{1}{2}\{1-(-1)^{i+j}\}Y_{i,j},
\end{equation}
anticommutes with the Hamiltonian (\ref{ham}). This means that for
every eigenvector $|v\rangle$ satisfying
${\cal H}(\alpha)|v\rangle=E(\alpha)|v\rangle$
we have another eigenvector $|w\rangle=S|v\rangle$ that satisfies
${\cal H}(\alpha)|w\rangle=-E(\alpha)|w\rangle$. This proves that for
even values of $L$ spectrum of ${\cal H}(\alpha)$ is symmetric around
zero but for odd $L$'s this property does not hold; then $S$ no longer
anticommutes with the Hamiltonian. To obtain a symmetric spectrum in
this case we would have to impose open boundary conditions. We would
like to emphasize that both Lanczos and KPM calculation for $6\times6$
lattice (with $2^{36}$-dimensional Hilbert space) would be nearly
impossible without using the symmetry operators and reduced
Hamiltonians given by Eqs. (\ref{Hiks}) and (\ref{Hzet}).

\begin{figure}[t!]
\begin{center}\includegraphics[width=8cm]{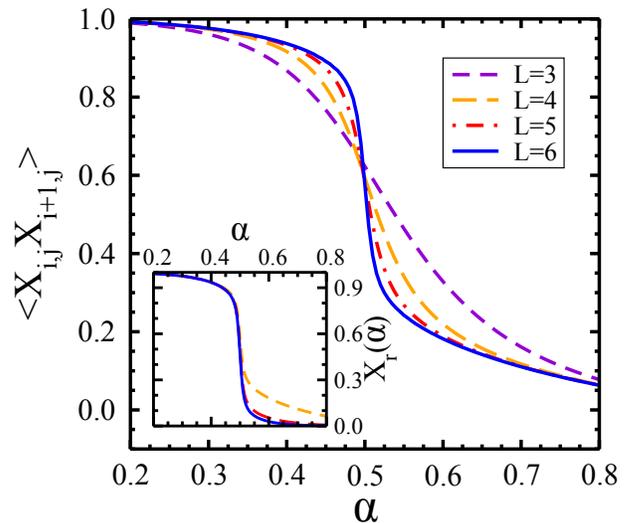}
\end{center}\caption{Nearest-neighbor spin correlations
$\langle X_{i,j}X_{i+1,j}\rangle$
for different cluster sizes $L$ and long range correlations (inset)
$X_{r}(\alpha)\equiv\langle X_{i,j}X_{i+r,j}\rangle$ for $L=6$ and
$r=1,2,3$ shown with dashed (orange), long dashed (red) and solid
(blue) lines, respectively. }
\label{fig:corxx}
\end{figure}

\subsection{Ground state properties}
\label{sub:gs}

In Fig. \ref{fig:corxx} we compare NN correlations
$\langle X_{i,j}X_{i+1,j}\rangle$ as functions of $\alpha$ obtained via
Lanczos algorithm for clusters of the sizes $L=3,4,5,6$. Curves for
finite clusters converge to certain final functions with an infinite
slope at $\alpha=\frac12$ but not to a step function which would mean
completely classical behavior. This result shows that even in the limit
of large $L$ the 2D QCM preserves quantum correction even though it
chooses to order in only one direction.\cite{Wen08} Looking at the inset
of Fig. \ref{fig:corxx} we can see longer-range correlations of the form
$X_{r}(\alpha)\equiv\langle X_{i,j}X_{i+r,j}\rangle$ (for symmetry
reasons any other two-point correlation functions involving $X_{i,j}$
operators must be zero in the ground state) for the $L=6$ system and
$r=1,2,3$. Their behavior is very similar to the NN correlations in the
sector of $\alpha\le\frac12$ but for $\alpha>\frac12$ they are
strongly suppressed and effectively behave in a more classical way.

\begin{figure}[t!]
\includegraphics[width=8cm]{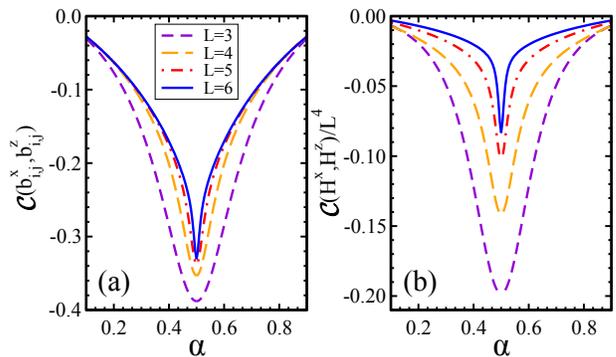}
\caption{Local and nonlocal ground state covariances of:
(a) the bond operators $b_{i,j}^{x}$ and $b_{i,j}^{z}$, and
(b) the $x$ and $z-$part of the Hamiltonian, $H^{x}$ and $H^{z}$,
normalized by $L^{4}$ for for different cluster sizes $L$
(note that the result for $L=3$ was scaled by factor $\frac15$). }
\label{fig:covs}
\end{figure}

In Fig. \ref{fig:covs}(a) we show the ground state covariances of the
bond operators $b_{i,j}^{x}\equiv X_{i,j}X_{i+1,j}$ and
$b_{i,j}^{z}\equiv Z_{i,j}Z_{i,j+1}$, i.e.,
\begin{equation}
{\cal C}(b_{i,j}^{x},b_{i,j}^{z})=
\left\langle b_{i,j}^{x}b_{i,j}^{z}\right\rangle
-\left\langle b_{i,j}^{x}\right\rangle\left\langle b_{i,j}^{z}\right\rangle.
\end{equation}
Analogous covariances for whole $x$ and $z-$part of the Hamiltonian,
namely $H^{x(z)}=\sum_{i,j}b_{i,j}^{x(z)}$, are shown in Fig.
\ref{fig:covs}(b) as normalized by the total number of terms in
$H^{x}H^{z}$, i.e., $L^{4}$. All covariances are of maximal magnitude at
$\alpha=\frac12$ and get suppressed when system size increases. In case
of bond covariances suppression is not total and we have some finite
covariance in whole range of $\alpha$ with cusp at $\alpha=\frac12$,
indicating singular behavior at this point. This means that locally
vertical and horizontal bonds cannot be factorized despite the fact that
the system chooses only one direction of ordering. On the other hand,
the normalized nonlocal covariance of $H^{x}$ and $H^{z}$ tends to
vanish for all $\alpha$ in the thermodynamic limit meaning that in the
long-range vertical and horizontal bonds behave as independent from one
another. We believe that this justifies the onset of directionally
ordered phase for $L\to\infty$.

Another interesting quantity that can be calculated in the ground state
is the von Neumann entropy of a chosen subsystem. This entropy tells us
to what extent the full wave function of the system cannot be factorized
and written as the wave function of the subsystem multiplied by the wave
function of the rest. In case of the QCM on square lattice the most
promising choice of a subsystem would be a single column or a single row
of the square lattice. To calculate von Neumann entropy of a column we
need to use its reduced density matrix $\rho_{L}$, defined as a partial
trace of a full density matrix,
\begin{equation}
\rho=\left|\Psi_{0}\right\rangle \left\langle \Psi_{0}\right|,
\label{rho}
\end{equation}
taken over the spins outside the column. This definition, however true,
is not very practical. For systems with spin $s=\frac12$ one can derive
a simpler formula:\cite{Eri09}
\begin{equation}
\rho_{L}=\frac{1}{2^{L}}\sum_{\mu_{1},..,\mu_{L}}\left\langle
\sigma_{1}^{\mu_{1}}...\sigma_{L}^{\mu_{L}}\right\rangle
\sigma_{1}^{\mu_{1}}...\sigma_{L}^{\mu_{L}}.
\label{eq:rhol}
\end{equation}
Here $\mu_{i}=0,x,y,z$, $\sigma_{i}^{0}=1$ and $\sigma_{i}^{\mu_{i}}$
are the spins taken from one column of a square cluster.

After diagonalizing $\rho_{L}$, which is of the size $2^{L}\times2^{L}$,
one can easily calculate von Neumann entropy as:
\begin{equation}
{\cal S}_{L}=-\mathrm{Tr}\rho_{L}\log_{2}\rho_{L}.
\label{S_L}
\end{equation}
For the symmetry reasons, described in detail in Sec. \ref{sec:sym},
Eq. (\ref{eq:rhol}) simplifies greatly as only the $x$-component spin
operators multiplied along the columns can give a finite average in the
ground state and their number must be even. Thus, for $L\le 6$ systems,
the matrix $\rho_L$ can be constructed with two--point, four--point and
single six--point correlation functions at most. Again, the reduced
form of the compass Hamiltonian simplifies getting the ground state but
we have to keep in mind that $\rho_L$ is expressed in terms of original
spins.

\begin{figure}[t!]
\includegraphics[width=8cm]{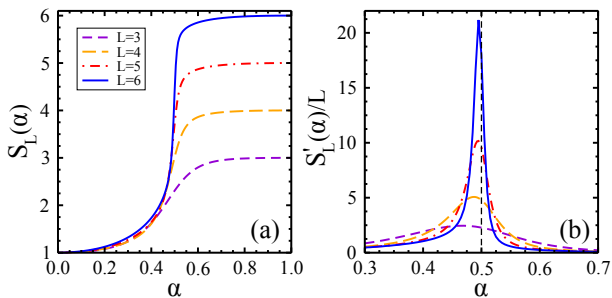}
\caption{
Panel (a) --- von Neumann entropy ${\cal S}_{L}(\alpha)$ (\ref{S_L})
of a column in the lattice of the size $L=3,4,5,6$ as a function of
$\alpha$.
Panel (b) --- derivative of von Neumann entropy $S{\cal S}_L(\alpha)$ with
respect to $\alpha$ normalized by $L$.
The line character as in panel (a).}
\label{cmp_ent}
\end{figure}

The results of von Neumann entropy calculations for a column of length
$L$ belonging to the $L\times L$ cluster is shown in Fig.
\ref{cmp_ent}(a). We can see that for $\alpha=0$ the entropy
$S_{L}(\alpha)$ is finite as we expect from the product state. On the
other hand, the system at $\alpha=0$ is purely classical and the
Hamiltonian (\ref{ham}) describes a set of noninteracting Ising
columns. Why is the ground state not a product of such columnar states?
This is not visible for the present choice of basis adapted for the
reduced form of the compass Hamiltonian given by Eqs. (\ref{Hiks}) and
(\ref{Hzet}).
Because of the spin transformations (\ref{transx}) and (\ref{transz}),
the ground state from the subspace $r_{i}\equiv s_{i}\equiv1$ found
here is a superposition of two-column product states with equal weights
and gives $S{\cal S}_{L}(0)=1$. This stays in agreement with the known
fact that von Neumann entropy \textit{depends} on the choice of basis
and this dependence comes precisely from the partial trace of the
density matrix $\rho$ Eq. (\ref{rho}). For that reason we should always
employ the most natural basis for a given problem.

For $0<\alpha<1$ the subspace $r_{i}\equiv s_{i}\equiv1$ is the most
natural one because it is the only subspace with the ground state (up to
global two--fold degeneracy). For $\alpha=0$ or $1$ the choice of the
eigenbasis of $\sigma_i^x$ or $\sigma_i^z$ operators seems to be more
natural one which implies ${\cal S}_{L}(0)={\cal S}_{L}(1)=0$ so the
plot in Fig. \ref{cmp_ent}(a) is valid only away from these points. The
upper limit for ${\cal S}_L(\alpha)$ is always $L$ which can be easily
proved by taking a state with all components equal. As we can see in
Fig. \ref{cmp_ent}(a) this limit is reached for $\alpha\to1$ and before
we have a region of abrupt change in ${\cal S}_{L}(\alpha)$ with slope
growing with increasing size $L$.
As a consequence, the derivative of ${\cal S}_L(\alpha)$ with respect to
$\alpha$ normalized by $L$ increases with system size, see Fig.
\ref{cmp_ent}(b). Because of the above normalization the area under the
plot is constant and equal to $1$. As we can see the curve tends to a
delta function centered around $\alpha=\frac12$ for growing system
size $L$. This suggests that there is a quantum phase transition of the
second order at $\alpha=\frac12$ in the thermodynamic limit because
second derivative of von Neumann entropy diverges, which stays in
analogy to the classical entropy and classical phase transition.

\begin{figure}[t!]
\includegraphics[width=8cm]{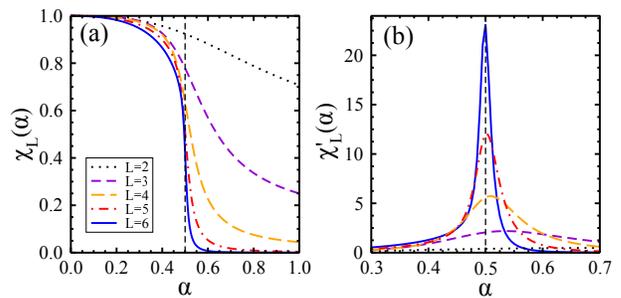}
\caption{Panel (a) --- fidelity $\chi_{L}(\alpha)$ (\ref{chi_L}) for
different values of $L=2,\dots,6$.
Panel (b) --- derivative of the fidelity $\chi_{L}(\alpha)$ with respect
to $\alpha$ for the same values of $\alpha$. }
\label{fig:cmp_ovr}
\end{figure}

We have also examined the overlap of the ground states obtained for
$\alpha$ to the left and to the right of (before and after) the
transition point at $\alpha=\frac12$ called also a {\it fidelity}. In
this case we are interested in fidelity $\chi_{L}(\alpha)$ defined as,
\begin{equation}
\chi_{L}(\alpha) \equiv \left\langle \psi_{0}(0^{+})\right.
\left|\psi_{0}(\alpha)\right\rangle ,
\label{chi_L}
\end{equation}
where $\left|\psi_{0}(\alpha)\right\rangle $ is a ground state for a
given $\alpha$ and $\left|\psi_{0}(0^{+})\right\rangle $ is one of two
possible $x$-ordered ground states for $\alpha$ being close to $0$. As
we can see from Fig. \ref{fig:cmp_ovr}(a), $\chi_{L}(\alpha)$ decays
monotonously for growing $\alpha$ and the drop is most pronounced around
$\alpha=\frac12$, especially for few largest $L$, as one could expect.
Less expected is that for small system sizes $\chi_{L}(\alpha)$ does not
vanish at $\alpha=1$ though the $x$-order changes completely to the
$z$-one. This effect is again due to the symmetries --- the symmetry
obeying ground state at $\alpha=0^{+}$ is a linear combination of the
classical configurations found at $\alpha=0$ and hence not necessarily
orthogonal to the one at $\alpha=1^{-}$. On the other hand, for growing
system size these states become more and more orthogonal and already
for $L=4$ we find that $\chi_{L}(\alpha)$ vanishes at $\alpha=1^{-}$.

In the extreme case of $L=6$ system $\chi_{L}(\alpha)$ is strongly
suppressed already at $\alpha=0.6$. Also in the $\alpha<\frac12$ regime
the fidelity drops faster for larger systems. The overall shape of the
limiting $\chi_{L=6}(\alpha)$ curve is very similar to the one obtained
with von Neumann entropy ${\cal S}_{L=6}(\alpha)$ of Fig.
\ref{cmp_ent}(a), especially below $\alpha=\frac12$, showing the
universality of the transition. Also the behaviors of the derivatives
are qualitatively the same, see Figs. \ref{fig:cmp_ovr}(b) and
\ref{cmp_ent}(b).

\subsection{The structure of energy levels for $L=6$}
\label{sub:ele}

The discrete energy spectrum of the Ising model changes into a dense
spectrum of the QCM when $\alpha$ increases, see Fig.
\ref{fig:fullspec}, where we show the results of full brute-force ED
of the $4\times4$ periodic cluster (this task is
impossible without using the symmetries of the QCM). All negative-energy
levels for $0\le\alpha\le\frac12$ are shown; full spectrum can be
constructed from the plot in Fig. \ref{fig:fullspec} by the mirror
reflections with respect to $\alpha=\frac12$ and $E_n=0$ axes. The
structure of energy levels undergoes the evolution from the ladder-like
classical excitation spectrum at $\alpha=0$ to the dense spectrum, with
low-energy states separated by small gaps and a quasi-continuum
structure at higher energy, at the isotropic point $\alpha=\frac12$.
Excited states at $\alpha=0$, being defected classical antiferromagnetic
chains with energy determined by the number of spin defects, are well
separated from one another until $\alpha\simeq 0.2$. For a fixed value
of $\alpha$, increasing energy reflects increasing number of spin
defects. The states with the lowest excitation
energy, corresponding to a single spin defect, are less susceptible to
the mixing caused by transverse terms in $H^z$, and remain separated
from other states almost until the transition point at $\alpha=\frac12$.
Even at this point the mixing involves only singly and doubly defected
states.

\begin{figure}[t!]
\caption{\label{fig:fullspec} All the energy levels $E_{n}$ of the $L=4$
system as functions of $\alpha\in[0,\frac12]$. The result is symmetric
with respect to $E_{n}=0$ and $\alpha=\frac12$ axes.
[This figure is not reproduced here for technical reasons;
it will appear in the published version of this paper.]}
\end{figure}

From the form of the QCM Hamiltonian (\ref{ham}), one can easily infer
the relation between the slope of the energy level $E_{n}$ and the
preferred ordering direction in the state $\left|\Psi_{n}\right\rangle $:
\begin{equation}
\frac{1}{J}\frac{d}{d\alpha}E_{n}(\alpha)=\left\langle
\Psi_{n}(\alpha)\right|H^{x}-H^{z}\left|\Psi_{n}(\alpha)\right\rangle ,
\end{equation}
which means that states ordered by $H^x$ are related with energy
levels with positive slope and the others are related with energy
levels with negative slope. Zero slope indicates that the state has
no preferred ordering direction; this happens to the ground state at
$\alpha=\frac12$ and the anticipated symmetry breaking between $H^{x}$
and $H^{z}$ implies that in the thermodynamic limit the lowest energy
level will have a cusp at this point because any infinitesimal deviation
from $\alpha=\frac12$ must lead to strictly positive or negative slope
of $E_{0}\left(\alpha=\frac12\pm\varepsilon\right)$
(as also shown by the PEPS simulations of Ref. \onlinecite{Oru09}).

Table \ref{tab:ene36_6} contains the two lowest energies, $E_0$ and
$E_1$, from each of the $36$ nonequivalent invariant subspaces of the
QCM for $L=6$ at $\alpha=\frac12$. Their degeneracies $d$ agree with the
considerations of Sec. \ref{sub:multi}. Table \ref{tab:ene36_6} however
was obtained by Lanczos recursions done in the full set of subspaces
and then the energies were compared to arrange the subspaces into the
classes. Similar Tables II-IV for the $L=2,3,4,5$ systems are presented
in the Appendix \ref{sec:tables}. Note that the $E_0$ energies appear in
the ascending order and that $E_0$'s from the first $8$ subspaces form a
multiplet of low lying states. This multiplet, already described in Ref.
\onlinecite{Mil05}, consists of the classical ground state
configurations at $\alpha=0$ and $\alpha=1$, split by the quantum
corrections at $\alpha=\frac12$. One could thus expect that the number
of states in the multiplet is equal to $2\times 2^{L}$ but it turns out
that the ground state of degeneracy $d=2$ is common for the two sets of
states so the multiplicity equals to $2\times2^{L}-2$. In case of the
$L=6$ system this gives $126$ and can be obtained by adding the
degeneracies of the first $8$ subspaces in Table \ref{tab:ene36_6}.
Looking at the results presented in the Appendix \ref{sec:tables}, one
can see that this holds for other system sizes as well.

\begin{table}[t!]
\caption{\label{tab:ene36_6} Ground state energy $E_{0}$ and first
excited state energy $E_{1}$ (both in the units of $J$) and their
degeneracies $d$ for $36$ nonequivalent subspaces of the $6\times6$
QCM Eq. (\ref{ham}) at $\alpha=\frac12$. States $n=1,\dots,8$
(bold face) come from the classical ground state manifolds at
$\alpha=0,1$ and their total number is $126=2\left(2^{6}\right)-2$. }
\centering{}%
\begin{ruledtabular}
\begin{tabular}{ccccccc}
 $n$  & \textbf{1} & \textbf{2} & \textbf{3} & \textbf{4} & \textbf{5}
& \textbf{6} \cr
\colrule
 $E_{0}$  & $-20.705$ & $-20.547$
& $-20.539$ & $-20.537$ & $-20.491$ & $-20.489$ \tabularnewline
$E_{1}$  & $-20.293$ & $-19.734$ & $-19.549$
& $-19.462$ & $-19.239$ & $-19.147$ \tabularnewline
 $d$  & 2 & 24 & 24 & 12 & 24 & 12 \tabularnewline
\hline
\hline
 $n$  & \textbf{7} & \textbf{8} & 9 & 10  & 11 & 12\tabularnewline
\colrule
 $E_{0}$  & $-20.489$ & $-20.451$ & $-20.050$ & $-19.984$
& $-19.965$ & $-19.877$\tabularnewline
$E_{1}$  & $-19.101$ & $-18.910$ & $-19.416$ & $-19.359$  & $-19.370$
& $-19.521$ \tabularnewline
 $d$  & 24 & 4 & 72 & 144 & 72 & 72\tabularnewline
\hline
\hline
 $n$  & 13 & 14 & 15 & 16 & 17 & 18\tabularnewline
\hline
 $E_{0}$  & $-19.835$ & $-19.834$ & $-19.814$ & $-19.813$
& $-19.722$ & $-19.707$\tabularnewline
$E_{1}$  & $-19.585$  & $-19.158$  & $-19.154$  & $-19.140$
& $-19.113$  & $-19.012$ \tabularnewline
 $d$  & 72 & 144 & 72 & 144 & 18 & 24\tabularnewline
\hline
\hline
 $n$  & 19 & 20 & 21 & 22 & 23 & 24\tabularnewline
\hline
 $E_{0}$  & $-19.675$  & $-19.627$ & $-19.622$ & $-19.611$
& $-19.522$ & $-19.461$\tabularnewline
$E_{1}$  & $-19.269$ & $-19.292$ & $-19.312$ & $-19.276$
& $-19.325$ & $-19.096$\tabularnewline
 $d$  & 144 & 72 & 144 & 72 & 36 & 24\tabularnewline
\hline
\hline
 $n$  & 25 & 26 & 27 & 28 & 29 & 30\tabularnewline
\hline
 $E_{0}$  & $-19.458$ & $-19.39$ & $-19.315$ & $-19.304$
& $-19.211$ & $-19.207$\tabularnewline
$E_{1}$  & $-19.151$ & $-18.869$ & $-18.850$ & $-18.880$
& $-19.036$ & $-18.741$\tabularnewline
 $d$  & 72 & 72 & 72 & 144 & 12 & 18\tabularnewline
\hline
\hline
 $n$  & 31 & 32 & 33 & 34 & 35 & 36\tabularnewline
 $E_{0}$  & $-19.175$ & $-19.073$ & $-19.068$ & $-18.900$
& $-18.714$ & $-18.264$\tabularnewline
$E_{1}$  & $-18.877$ & $-18.689$ & $-18.561$ & $-18.429$
& $-18.463$ & $-17.918$\tabularnewline
 $d$  & 72 & 72 & 24 & 12 & 24 & 2\tabularnewline
\end{tabular}
\end{ruledtabular}
\end{table}

\begin{figure}[t!]
\begin{center}
\includegraphics[width=7.5cm]{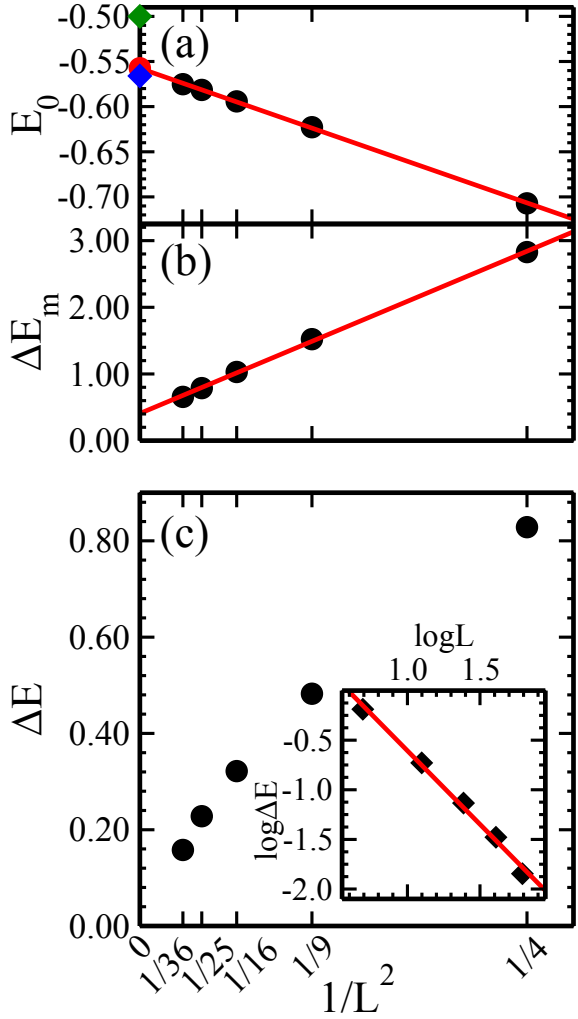}
\end{center}
\caption{Extrapolations in $1/L^{2}$ for the $\alpha=\frac12$ QCM.
Filled circles represent: (a) the ground state energy $\varepsilon_0$
per site; (b) the gap $\Delta E_m$ between the ground state multiplet
and the higher lying states; and (c) the energy gap $\Delta E$ to the
first excited state. The (red) lines are the linear fits to the data for
panels (a), (b). The inset of panel (c) shows the log-log plot for the
energy gap $\Delta E$. The
(red) dots at $1/L^2=0$ are the values predicted by the fits and square
and diamond of panel (a) are the classical $\varepsilon_0^{\rm clas}$
and CMF $\varepsilon_{0}^{\rm CMF}$ extrapolated values of
$\varepsilon_0$.
}
\label{fig:extr}
\end{figure}

In Fig. \ref{fig:extr} we show the extrapolation of different energies
for the infinite system done using the data from Table \ref{tab:ene36_6}
and from the Appendix. Fig. \ref{fig:extr}(a) shows the behavior of the
ground state energy per site $\varepsilon_0$ as a function of $1/L^{2}$.
As we can see the data points nicely lie on a straight line and the
linear fit gives the extrapolated ground state energy per site equal to
\begin{equation}
\varepsilon_{0}(L\to\infty)=-(0.5575\pm 0.0007)J,
\label{ex}
\end{equation}
This value lies between
the classical value of $\varepsilon_{0}^{\rm clas}=-0.5 J$ which
one can get by keeping only one part of the Hamiltonian, either $H^x$
or $H^z$, and a chain MF (CMF) value,
$\varepsilon_0^{\rm CMF}\simeq -0.5661J$.
The CMF approach that we used here relies on splitting the interaction
along one direction and treating the system as a set of Ising chains in
a transverse field coupled to the MF (see Ref. \onlinecite{wb11} for
more details). Such chains can be then solved exactly so the quantum
physics within a single chain is well captured. Being variational, the
CMF approach must give higher ground state energy than the exact ground
state energy, and indeed the PEPS estimation for $\varepsilon_0$ is
$\varepsilon_0^{\rm PEPS}\simeq -0.5684J$ (originally for different
parametrization of interactions, see Ref. \onlinecite{Oru09}), is only
slightly lower than $\varepsilon_0^{\rm CMF}$. Therefore we can conclude
that the linear extrapolation (\ref{ex}) is not fully satisfactory
though all three energies lie indeed close to one another. Surprisingly,
the CMF description of the 2D QCM turns out to be quite precise.

Fig. \ref{fig:extr}(b) depicts the energy gap $\Delta E_{m}$ between the
ground state energy of the last subspace from the ground state multiplet
and the lowest energy of the remaining subspaces. For instance, in case
of the $L=6$ system the gap reads $\Delta E_m=E_0(n=9)-E_0(n=8)$. Like
before, this quantity shows relatively good linear behavior as a
function of $1/L^2$ (however small negative curvature can be observed
for $L=6$) and the extrapolation in $L\to\infty$ can be easily performed
to obtain $\Delta E_m(L\to\infty)=(0.408\pm 0.018)J$. The finite value
of this gap for $L\to\infty$ means that the spectrum of the system
divides into the low lying set of states separated by energy of
$\Delta E_m$ from the rest.
We argue that these states are the space of all possible
nematic-ordered states and the finite value of $\Delta E_m$ makes the
order robust at finite temperature up to $T_{c}=0.055J$ as shown in Ref.
\onlinecite{Wen08}. For the complete picture one could show that the
width of the multiplet tends to zero for $L\to\infty$ but unfortunately
its behavior as function of $1/L^2$ is quite irregular in this range of
$L$.

Last but not least, we present the energy gap $\Delta E$ between the
ground and the first excited state as a function of $1/L^2$ in Fig.
\ref{fig:extr}(c). Note that this gap is a different quantity than the
gaps discussed in Ref. \onlinecite{Mil05} and is equivalent to the
excitation energy associated with flipping one of the classical spins
$\{r_i,s_i\}$ of the reduced Hamiltonian of Eqs. (\ref{Hiks}) and
(\ref{Hzet}). Surprisingly, it turns out that the gap does not
decay exponentially with $L^2$ or $L$ (as it happens for the 1D
transverse-field Ising model) but exhibits rather a power-law behavior.
This can be seen more easily on a log-log plot in the inset of Fig.
\ref{fig:extr}(c) where the data points show quite good linear behavior.
The power-law fit of the form $\Delta E\propto(1/L^2)^{1/\chi}$ gives
critical exponent $\chi=1.418\pm0.043$ which can be related to the
dynamical critical exponent $z$ as $z=2/\chi$ (for the imaginary-time
dynamical correlation length $\xi_{\tau}$ behaves like
$\xi_{\tau}\propto 1/\Delta E$ and at the critical point
$\xi_{\tau}\propto\xi^{z}$, see Ref. \onlinecite{Par10}).
Thus finally we obtain $z=1.409\pm 0.042$.

\subsection{Density of states and specific heat}
\label{sub:cv}

The main benefit for ED calculations is that after the transformation
the Hamiltonian of $L\times L$ compass model ($\alpha=\frac12$) turns
into $2^{2L-1}$ spin models, each one on an $(L-1)\times(L-1)$ lattice.
In fact, the number of \textit{different} models is much lower than
$2^{2L-1}$; most of the resulting Hamiltonians differ only by a
similarity transformation as shown in the Sec. \ref{sub:multi}. For
example, in case of the $6\times6$ system we find out that only $36$ out
of $2048$ Hamiltonians are different; their two lowest energies obtained
using the Lanczos algorithm, and their degeneracies are given in Table
\ref{tab:ene36_6}. Similar data for lower system sizes can be found in
the Appendix (in fact, these energies are known with much higher
precision and up to $L=5$ we have determined all the high energy states).

\begin{figure}[t!]
\includegraphics[width=8cm]{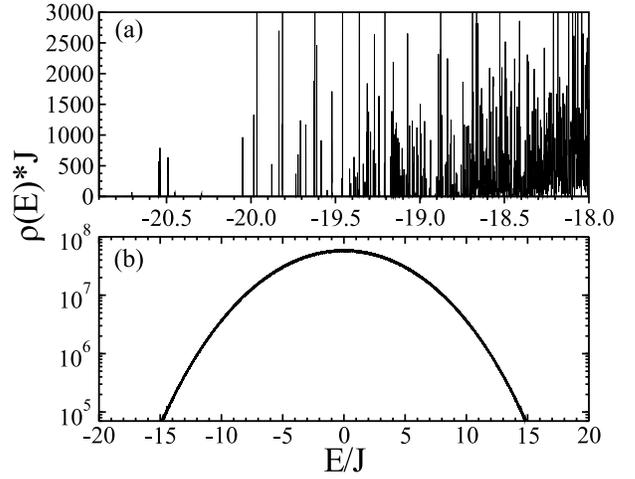}
\caption{Density of states $\rho(E)$ for the $6\times6$ compass cluster
at $\alpha=\frac12$. Panel (a) --- low energy region, lowest lying
peaks agree with results of Lanczos recursion, excitation spectrum
is discrete; panel (b) --- full energy range in the logarithmic scale,
parabolic behavior indicates dense gaussian spectrum of high-energy
excitations. }
\label{fig:rho}
\end{figure}

This brings us to the calculation method --- the KPM
based on the expansion into the series of
Chebyshev polynomials.\cite{Feh08} Chebyshev polynomial of the $n$-th
degree is defined as $T_{n}(x)=\cos[n\arccos x]$ where $x\in[-1,1]$ and
$n$ is integer. Further on, we are going to calculate $T_{n}$ of the
Hamiltonian so first we need to renormalize it so that its spectrum fits
into the interval $[-1,1]$. This can be done easily if we know the width
of the spectrum. Our aim is to calculate the renormalized density of
states $\tilde{\rho}(E)$ given by
\begin{equation}
\tilde{\rho}(E)=(1/D)\sum_{n=0}^{D-1}\delta(E-\tilde{E_{n}}),
\end{equation}
where the sum is over eigenstates of ${\cal H}(\alpha)$ and $D$ is the
dimension of the Hilbert space. The moments $\mu_{n}$ of the expansion of
$\tilde{\rho}(E)$ in basis of Chebyshev polynomials can be expressed by:
\begin{equation}
\mu_{n}=\int_{-1}^{1}T_{n}(E)\tilde{\rho}(E)dE=
\frac{1}{D}\textrm{Tr}\{T_{n}(\tilde{{\cal H}})\}\,.
\end{equation}
Trace can be efficiently estimated using stochastic approximation:
\begin{equation}
\textrm{Tr}\,\{T_{n}(\tilde{{\cal H}})\}\approx\frac{1}{R}\sum_{r=1}^{R}
\;\langle r|T_{n}(\tilde{{\cal H}})|r\rangle\,,
\end{equation}
where $|r\rangle$ ($r=1,2,\dots,R$) are randomly picked complex
vectors with components $\chi_{r,k}$ ($k=1,2,\dots,D$) satisfying
$\langle\chi_{r,k}\rangle=0$, $\langle\chi_{r,k}\chi_{r',l}\rangle=0$,
$\langle\bar{\chi}_{r,k}\chi_{r',l}\rangle=\delta_{r,r'}\delta_{k,l}$
(the average is taken over the probability distribution). This
approximation converges very rapidly to the true value of the trace,
especially for a large value of $D$.

Action of the $T_{n}(\tilde{{\cal H}})$ operator on a vector $|r\rangle$
can be determined recursively using the following relation between
Chebyshev polynomials:
\begin{equation}
T_{n}(\tilde{{\cal H}})|r\rangle=
\{2\tilde{{\cal H}}\, T_{n-1}(\tilde{{\cal H}})-T_{n-2}
(\tilde{{\cal H}})\}|r\rangle.
\end{equation}
We can also use the relation
\begin{equation}
2T_{m}(x)T_{n}(x)=T_{m+n}(x)+T_{m-n}(x)
\end{equation}
to get moments $\mu_{2n}$ from the polynomials of the degree $n$.
Finally, the required function,
\begin{equation}
\tilde{\rho}(E)\approx\frac{1}{\pi\sqrt{1-E^{2}}}\left\{
g_{0}\mu_{0}+2\sum_{n=1}^{N-1}g_{n}\mu_{n}T_{n}(E)\right\} \,,
\end{equation}
can be reconstructed from the $N$ known moments, where coefficients
$\{g_{n}\}$ come from the integral kernel we use for better
convergence. Here we use Jackson kernel. Choosing the arguments of
$\tilde{\rho}(E)$ as being equal to $E_{k}=\cos[(2k-1)\pi/2N']$
($k=1,2,\dots,N'$) we can change the last formula into a cosine Fourier
series and use fast Fourier transform algorithms to obtain
$\tilde{\rho}(E_{k})$ rapidly. This point is crucial when $N$ and $N'$
are large, which is the case here; our choice will be $N=20000$ and
$N'=2N$. Using this procedure we can get the density of states for
$L=4,5,6$ systems. After getting energy spectra for the nonequivalent
subspaces we sum them with proper degeneracy factors to get the final
density of states $\tilde{\rho}(E)$ and next the partition function via
rescaling and numerical integration.

\begin{figure}[t!]
\includegraphics[width=8cm]{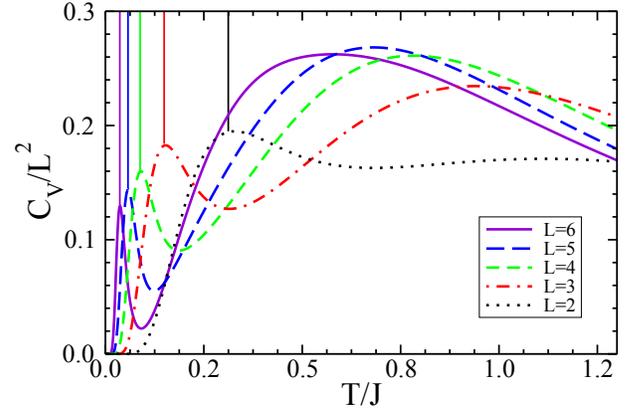}
\caption{Specific heats $C_{V}/L^{2}$ per site as functions of temperature
$T/J$ for $\alpha=\frac12$ obtained for the QCM clusters of increasing size
$L=2,\dots,6$. Vertical lines show the position of the low-energy peak in
$C_V$ according to the canonical ensemble over the ground state multiplet
only. }
\label{fig:heat}
\end{figure}

In Figs. \ref{fig:rho}(a) and \ref{fig:rho}(b) we display the density
of states $\rho(E)$ (without normalization) for the system size $L=6$.
Achieved resolution is such that one can distinguish single low-lying
energy states and the positions of peaks agree with the results of
Lanczos algorithm, see panel (a). In addition we get information about
the degeneracy of energy levels encoded in the area below the peaks.
This required very time-consuming calculations as the size of Hilbert
space is above $30$ million. In Fig. \ref{fig:rho}(b) we present an
overall view of full density of states in the logarithmic scale
exhibiting gaussian behavior. Note different orders of magnitude in
Figs. \ref{fig:rho}(a) and \ref{fig:rho}(b).
Both plots show that in the thermodynamic limit the spectrum of the 2D
QCM can be discrete in the lowest and highest-energy region and
continuous in the center, which agrees with
the existence of ordered phase above $T=0$.\cite{Wen08}

\begin{figure}[t!]
\begin{center}
\includegraphics[width=7.5cm]{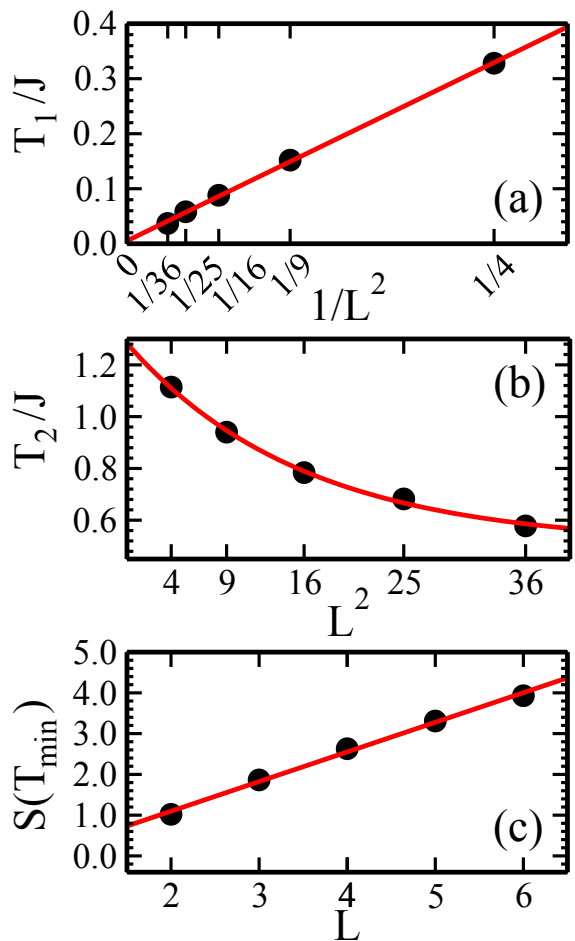}
\end{center}
\caption{Finite-size extrapolations related with specific heat curves of
Fig. \ref{fig:heat}: (a) --- position $T_1$ of the low-temperature peak,
(b) --- position $T_2$ of the high-temperature peak, and (c) --- entropy
$S(T_{\rm min})$ at the temperature $T_{\rm min}$ of the dip between the
two peaks in specific heat. Lines are the fits to the data points
(filled circles).}
\label{fig:extr2}
\end{figure}

In Fig. \ref{fig:heat} we show the specific heat $C_{V}/L^{2}$ obtained
for the compass $L\times L$ clusters calculated from:
(i) the densities of states $\rho(E)$ for $L=5,6$, and
(ii) the full energy spectrum for $L=2,3,4$.
Additionally, to enhance the precision at low temperatures the
lowest-lying energies obtained via stabilized Lanczos algorithm were
used up to certain energy above $E_{0}$. For $L=5$ system all the
energies up to $E\approx E_{0}+2J$ were determined by Lanczos algorithm
but for $L=6$ only a few states above $E_{0}$ could be found due to the
large size of the Hilbert space. The curves of specific heat of Fig.
\ref{fig:heat} exhibit two-peak structure similar to the one observed
for a compass ladder (see Ref. \onlinecite{wb10}), but in contrary to the
ladder case the low-temperature peak seems to vanish for $L\to\infty$ and
the specific heat develops a gap before the high-temperature peak.

On the other hand, we can see that the position of the low-temperature
peak agrees well with multiplet structure of the low-lying energy
levels described in Sec. \ref{sub:ele} --- one can calculate the
partition function over them to obtain the low-energy specific heat,
then we can determine the position of its peak and compare it with the
plot of Fig. \ref{fig:heat}. As we can see the small peak coincides with
the ground state multiplet peak for all values of $L$.

In Figs. \ref{fig:extr2}(a) and \ref{fig:extr2}(b) we show the positions
$T_1$ and $T_2$ of the low- and high-temperature peaks as functions of
$1/L^2$ and $L^2$, respectively. As we can see from its linear behavior
$T_1$ scales as $1/L^2$ and its extrapolated value for $L\to\infty$ is
zero. On the other hand, the linear fit for $T_2$ as function of
$1/L^2$ turns out to be unsatisfactory and the best obtained fit is
of the exponential form, with
\begin{equation}
T_2(L\to\infty)=(0.505\pm 0.035)J.
\end{equation}
Finally, in Fig. \ref{fig:extr2}(c) we show the scaling behavior of the
entropy $S(T_{\rm min})$ calculated from the specific heat at the
temperature $T_{\rm min}$ being the minimum between the two peaks in
$C_{V}$ which separates the low- and the high-energy excitations of the
model, see Fig. \ref{fig:heat}. As one could expect $S(T_{\rm min})$
scales linearly in $L$ because the number of the low-lying states is
of the order of $2^L$ as shown in Section \ref{sub:ele}.

Effectively, the present data suggest that the specific heat curve in
the thermodynamic limit would be rather like the one of classical Ising
ladder (see Ref. \onlinecite{wb10}), with a single broad peak in the
high temperature regime and zero specific heat up to certain $T_{0}$,
than the one of a compass ladder with robust low-energy excitations.
This means that the thermal behavior of the 2D QCM is indeed mostly
classical and agrees with the presence of ordered phase for finite $T$
in the thermodynamic limit.\cite{Wen08}

\section{Summary and conclusions}
\label{sec:summa}

We have presented the consequences of symmetry properties of the 2D
QCM which is in the center of interest at present.\cite{vdB13}
Using this example we argue that for a certain class of pseudospin
models, which have lower symmetry than SU(2), the spectral properties
can be uniquely determined by discrete symmetries like parity. In the
case of the conservation of spin parities in rows and columns in the 2D
QCM (for $x$ and $z$-components of spins), we have observed that the
ground state behaves according to a nonlocal Hamiltonian Eqs.
(\ref{Hiks}) and (\ref{Hzet}). In the ground state most of the
two-site spin correlations vanish and the two-dimer correlations exhibit
the nontrivial hidden order. For a finite system, the low-energy
excitations are the ground states of the QCM Hamiltonians in different
invariant subspaces which, as shown for the $Q_j$ symmetries,\cite{Mil05}
become degenerate with the ground state in the thermodynamic limit,
leading to degeneracy $d$ being exponential in the linear system size
$L$ ($d=2^{2L-1}$ or larger, if one could count the excited states from
different subspaces). The invariant subspaces can be classified by
lattice translations --- the reduction of the Hilbert space achieved in
this way is important for future numerical studies of the QCM and will
play a role for spin models with similar symmetries.

The reduced QCM Hamiltonian turned out to be very useful for the
state-of-the-art implementations of the ED techniques and gives the
access to the system sizes unavailable otherwise. In contrast to the
point-group or translational symmetries often explored for such models,
spin transformations lead to spin Hamiltonian again which makes it
particularly easy to implement. Although QCM has no sign problem and can
be treated with powerful quantum Monte Carlo methods, ED gives most
complete solution: the ground state wave function giving the access to
all possible correlators and measures of entanglement. Using Lanczos and
full diagonalization techniques we showed the behavior of all two-point
correlation functions for different system sizes and the full structure
of energy levels as functions of anisotropy parameter $\alpha$,
indication of discrete-continuum nature of the spectrum of the QCM.
Finally, we have obtained the ground state energy $\varepsilon_0$
per site up to $L=6$ and its extrapolation in the limit of $L\to\infty$,
$\varepsilon_{0}(L\to\infty)=-(0.5575\pm 0.0007)J$,
which is very close indeed to the CMF result,
$\varepsilon_0^{\rm CMF}\approx -0.5661J$. Both values are also very
close to the best estimate known from PEPS for the QCM,\cite{Oru09}
$\varepsilon_0^{\rm PEPS}\approx -0.5684J$.

The behavior of von Neumann entropy ${\cal S}_L(\alpha)$ of a single
column of a square lattice together with the fidelity $\chi_L(\alpha)$
and the energy gap $\Delta E$ at $\alpha=\frac12$, which decays in a
power-low fashion for growing $L$, suggests that the phase transition at
$\alpha=\frac12$ is of the second order with dynamical critical exponent
$z=1.409\pm0.042$. On the other hand, there is strong evidence, provided
by the PEPS simulations,\cite{Oru09} that the transition is indeed of
the first order. Following the idea of Ref. \onlinecite{Eri09} we argue
that this discrepancy could be cured by adopting a similar scenario of
a phase transition to the one suggested for the 1D QCM (see Ref.
\onlinecite{Eri09}) --- $\alpha=\frac12$ could be a multicritical point
of a more general model whose special case is the isotropic QCM
thus the transition carries the features of both first and second order.

Summarizing, using Kernel Polynomial Method we have gained the access to
the full density of states function $\rho(E)$ for system sizes excluding
full ED, i.e., for $L=5,6$. The obtained $\rho(E)$ for $L=6$ confirms
that the spectrum consists of discrete states at low energy, accompanied
by the continuum part at higher energy, as observed before for smaller
system size ($L=4$).\cite{Mil05} In addition, the extrapolation of the
gap $\Delta E_m$ in the limit $L\to\infty$ shows that the manifold of
the low-lying states, which collapse to the degenerate ground state in
the $L\to\infty$ limit,\cite{Mil05} develops a gap to the higher-lying
states of the width $\Delta E_m=(0.408\pm 0.018)J$. This supports the
existence of an ordered quasi-1D nematic phase at finite temperature.

It is quite remarkable that the specific heat of the system, calculated
from $\rho(E)$ for growing $L$, evolves to the curve characteristic
for a classical Ising ladder,\cite{wb10} with a single broad peak at
$T_2=(0.505\pm 0.035)J$ and a gap in low temperature, as shown by the
finite-size extrapolation. Within the error bar this is half of the
classical excitation energy when only interactions along a single
direction contribute. While the specific heat for a finite system
consists of two characteristic peaks, we demonstrated a distinct
behavior of these peaks:
(i) the position of the broad peak at high temperature saturates
exponentially with increasing system size $L^2$, and
(ii) the low-temperature maximum decreases and its position
approaches zero as $1/L^2$.
Finally, the entropy related with the low-energy sector scales linearly
with $L$ which agrees with the number of states in the low-energy
manifold being of the order of $2^L$, as indicated before\cite{Mil05}
and confirmed by our analysis. This is another manifestation of a
classical behavior of the QCM at finite temperature.

\acknowledgments
We kindly acknowledge financial support by the Polish National Science
Center (NCN) under Project No. 2012/04/A/ST3/00331.

\appendix*

\section{Square clusters with $L<6$}
\label{sec:tables}

As a supplement to Table \ref{tab:ene36_6} of Sec. \ref{sub:ele},
we present here analogous Tables with energies and degeneracies for
inequivalent subspaces for other $L\times L$ clusters with $L<6$:
Table \ref{tab:ene10_5} for $L=5$, Table \ref{tab:ene10_4} for $L=4$,
and Table \ref{tab:ene3_3} for $L=3$ and for $L=2$.

\begin{table}[h!]
\caption{\label{tab:ene10_5} Ground state energy $E_0$ and first excited
state energy $E_{1}$ (in the units of $J$) and their degeneracies
$d$ for $10$ nonequivalent subspaces of the $5\times 5$ QCM Eq.
(\ref{ham}) at $\alpha=\frac12$. States $n=1,\dots,4$ (bold face)
come from the classical ground state manifolds at $\alpha=0,1$ and their
total number is $62=2\left(2^{5}\right)-2$. }
\begin{ruledtabular}
\begin{tabular}{cccccc}
 $n$  & \textbf{1} & \textbf{2} & \textbf{3} & \textbf{4} & \textbf{5}\cr
\hline
 $E_{0}$  & $-14.54$ & $-14.31$ & $-14.30$ & $-14.22$ & $-13.75$ \cr
 $E_{1}$  & $-13.80$ & $-13.15$ & $-12.91$ & $-12.50$ & $-12.86$ \cr
 $d$  & 2 & 20 & 20 & 20 & 50\cr
\hline
\hline
 $n$  & 6 & 7 & 8 & 9 & 10\cr
\hline
$E_{0}$  & $-13.67$ & $-13.52$ & $-13.45$ & $-13.22$ & $-12.79$ \cr
$E_{1}$  & $-12.99$ & $-13.26$ & $-12.67$ & $-12.88$ & $-12.30$ \cr
$d$  & 100 & 50 & 100 & 100 & 50 \cr
\end{tabular}
\end{ruledtabular}
\end{table}

\begin{table}[h!]
\caption{\label{tab:ene10_4} Ground state energy $E_0$ and first excited
state energy $E_{1}$ (in the units of $J$) and their degeneracies
$d$ for $10$ nonequivalent subspaces of the $4\times4$ QCM Eq.
(\ref{ham}) at $\alpha=1/2$. States $n=1,\dots,4$ (bold face)
come from the classical ground state manifolds at $\alpha=0,1$ and their
total number is $30=2\left(2^{4}\right)-2$. }
\begin{ruledtabular}
\begin{tabular}{cccccc}
 $n$  & \textbf{1} & \textbf{2} & \textbf{3} & \textbf{4} & \textbf{5}\cr
\hline
 $E_{0}$ & $-9.51$ & $-9.18$ & $-9.17$ & $-9.04$ & $-8.48$ \cr
 $E_{1}$ & $-8.17$ & $-7.32$ & $-7.36$ & $-6.76$ & $-7.46$ \cr
 $d$  & 2 & 16 & 8 & 4 & 32 \cr
\hline
\hline
 $n$  & 6 & 7 & 8 & 9 & 10 \cr
\hline
 $E_{0}$ & $-8.38$ & $-8.11$ & $-8.05$ & $-7.61$ & $-6.84$ \cr
 $E_{1}$ & $-7.69$ & $-6.98$ & $-7.12$ & $-7.33$ & $-6.48$ \cr
 $d$  & 32 & 8 & 16 & 8 & 2\cr
\end{tabular}
\end{ruledtabular}
\end{table}

\begin{table}[h!]
\caption{\label{tab:ene3_3} Ground state energy $E_0$ and first excited
state energy $E_1$ (in the units of $J$) and their degeneracies
$d$ for $3$ nonequivalent subspaces of the $3\times 3$ ($L=3$) and
$2\times2$ ($L=2$) QCM Eq.
(\ref{ham}) at $\alpha=\frac12$. States $n=1,2$ (bold face) come from
the classical ground state manifolds at $\alpha=0,1$ and their total
number is $2\left(2^{L}\right)-2$, i.e., 14 for $L=3$ and 6 for $L=2$. }
\begin{ruledtabular}
\begin{tabular}{ccccccc}
      & & $L=3$ &  & & $L=2$ &  \cr
 $n$  & \textbf{1} & \textbf{2} & 3 & \textbf{1} & \textbf{2} & 3 \cr
\hline
 $E_{0}$ & $-5.61$ & $-5.12$ & $-4.08$ & $-2.83$ & $-2.00$ & $0.00$ \cr
 $E_{1}$ & $-3.00$ & $-2.34$ & $-3.20$ & $+2.83$ & $+2.00$ & $0.00$ \cr
 $d$  & 2 & 12 & 18 & 2 & 4 & 2 \cr
\end{tabular}
\end{ruledtabular}
\end{table}

By comparing the data in Tables II-IV for different system size $L$,
we observe that the total width of the spectrum increases with
increasing $L$, but the first excitation energy $E_1-E_0$ decreases.
It is also remarkable that the number of nonequivalent subspaces in
the range of $L<6$ increases from odd $L$ to even $(L+1)$ but stays
constant from an even $L$ to the next odd $(L+1)$ size. No general
proof of this property could be found so far.

\vfill
\eject

\bibliographystyle{apsrev4-1}

\begin{thebibliography}{51}%
\makeatletter
\providecommand \@ifxundefined [1]{%
 \@ifx{#1\undefined}
}%
\providecommand \@ifnum [1]{%
 \ifnum #1\expandafter \@firstoftwo
 \else \expandafter \@secondoftwo
 \fi
}%
\providecommand \@ifx [1]{%
 \ifx #1\expandafter \@firstoftwo
 \else \expandafter \@secondoftwo
 \fi
}%
\providecommand \natexlab [1]{#1}%
\providecommand \enquote  [1]{``#1''}%
\providecommand \bibnamefont  [1]{#1}%
\providecommand \bibfnamefont [1]{#1}%
\providecommand \citenamefont [1]{#1}%
\providecommand \href@noop [0]{\@secondoftwo}%
\providecommand \href [0]{\begingroup \@sanitize@url \@href}%
\providecommand \@href[1]{\@@startlink{#1}\@@href}%
\providecommand \@@href[1]{\endgroup#1\@@endlink}%
\providecommand \@sanitize@url [0]{\catcode `\\12\catcode `\$12\catcode
  `\&12\catcode `\#12\catcode `\^12\catcode `\_12\catcode `\%12\relax}%
\providecommand \@@startlink[1]{}%
\providecommand \@@endlink[0]{}%
\providecommand \url  [0]{\begingroup\@sanitize@url \@url }%
\providecommand \@url [1]{\endgroup\@href {#1}{\urlprefix }}%
\providecommand \urlprefix  [0]{URL }%
\providecommand \Eprint [0]{\href }%
\providecommand \doibase [0]{http://dx.doi.org/}%
\providecommand \selectlanguage [0]{\@gobble}%
\providecommand \bibinfo  [0]{\@secondoftwo}%
\providecommand \bibfield  [0]{\@secondoftwo}%
\providecommand \translation [1]{[#1]}%
\providecommand \BibitemOpen [0]{}%
\providecommand \bibitemStop [0]{}%
\providecommand \bibitemNoStop [0]{.\EOS\space}%
\providecommand \EOS [0]{\spacefactor3000\relax}%
\providecommand \BibitemShut  [1]{\csname bibitem#1\endcsname}%
\let\auto@bib@innerbib\@empty
\bibitem [{\citenamefont {Tokura}\ and\ \citenamefont {Nagaosa}(2000)}]{Nag00}%
  \BibitemOpen
  \bibfield  {author} {\bibinfo {author} {\bibfnamefont {Y.}~\bibnamefont
  {Tokura}}\ and\ \bibinfo {author} {\bibfnamefont {N.}~\bibnamefont
  {Nagaosa}},\ }\href@noop {} {\bibfield  {journal} {\bibinfo  {journal}
  {Science}\ }\textbf {\bibinfo {volume} {288}},\ \bibinfo {pages} {462}
  (\bibinfo {year} {2000})}\BibitemShut {NoStop}%
\bibitem [{\citenamefont {Ole\'s}\ \emph {et~al.}(2005)\citenamefont {Ole\'s},
  \citenamefont {Khaliullin}, \citenamefont {Horsch},\ and\ \citenamefont
  {Feiner}}]{Ole05}%
  \BibitemOpen
  \bibfield  {author} {\bibinfo {author} {\bibfnamefont {A.~M.}\ \bibnamefont
  {Ole\'s}}, \bibinfo {author} {\bibfnamefont {G.}~\bibnamefont {Khaliullin}},
  \bibinfo {author} {\bibfnamefont {P.}~\bibnamefont {Horsch}}, \ and\ \bibinfo
  {author} {\bibfnamefont {L.~F.}\ \bibnamefont {Feiner}},\ }\href@noop {}
  {\bibfield  {journal} {\bibinfo  {journal} {Phys. Rev. B}\ }\textbf {\bibinfo
  {volume} {72}},\ \bibinfo {pages} {214431} (\bibinfo {year}
  {2005})}\BibitemShut {NoStop}%
\bibitem [{\citenamefont {Khaliullin}(2005)}]{Kha05}%
  \BibitemOpen
  \bibfield  {author} {\bibinfo {author} {\bibfnamefont {G.}~\bibnamefont
  {Khaliullin}},\ }\href@noop {} {\bibfield  {journal} {\bibinfo  {journal}
  {Prog. Theor. Phys. Suppl.}\ }\textbf {\bibinfo {volume} {160}} (\bibinfo
  {year} {2005})}\BibitemShut {NoStop}%
\bibitem [{\citenamefont {Ole\'s}(2012)}]{Ole12}%
  \BibitemOpen
  \bibfield  {author} {\bibinfo {author} {\bibfnamefont {A.~M.}\ \bibnamefont
  {Ole\'s}},\ }\href@noop {} {\bibfield  {journal} {\bibinfo  {journal} {J.
  Phys.: Condens. Matter}\ }\textbf {\bibinfo {volume} {24}},\ \bibinfo {pages}
  {313201} (\bibinfo {year} {2012})}\BibitemShut {NoStop}%
\bibitem [{\citenamefont {Feiner}\ \emph {et~al.}(1997)\citenamefont {Feiner},
  \citenamefont {Ole\'s},\ and\ \citenamefont {Zaanen}}]{Fei97}%
  \BibitemOpen
  \bibfield  {author} {\bibinfo {author} {\bibfnamefont {L.~F.}\ \bibnamefont
  {Feiner}}, \bibinfo {author} {\bibfnamefont {A.~M.}\ \bibnamefont {Ole\'s}},
  \ and\ \bibinfo {author} {\bibfnamefont {J.}~\bibnamefont {Zaanen}},\
  }\href@noop {} {\bibfield  {journal} {\bibinfo  {journal} {Phys. Rev. Lett.}\
  }\textbf {\bibinfo {volume} {78}},\ \bibinfo {pages} {2799} (\bibinfo {year}
  {1997})}\BibitemShut {NoStop}%
\bibitem [{\citenamefont {Feiner}\ \emph {et~al.}(1998)\citenamefont {Feiner},
  \citenamefont {Ole\'s},\ and\ \citenamefont {Zaanen}}]{Fei98}%
  \BibitemOpen
  \bibfield  {author} {\bibinfo {author} {\bibfnamefont {L.~F.}\ \bibnamefont
  {Feiner}}, \bibinfo {author} {\bibfnamefont {A.~M.}\ \bibnamefont {Ole\'s}},
  \ and\ \bibinfo {author} {\bibfnamefont {J.}~\bibnamefont {Zaanen}},\
  }\href@noop {} {\bibfield  {journal} {\bibinfo  {journal} {J. Phys.: Condens.
  Matter}\ }\textbf {\bibinfo {volume} {10}},\ \bibinfo {pages} {L555}
  (\bibinfo {year} {1998})}\BibitemShut {NoStop}%
\bibitem [{\citenamefont {Brzezicki}\ \emph {et~al.}(2012)\citenamefont
  {Brzezicki}, \citenamefont {Dziarmaga},\ and\ \citenamefont
  {Ole\'s}}]{Brz12}%
  \BibitemOpen
  \bibfield  {author} {\bibinfo {author} {\bibfnamefont {W.}~\bibnamefont
  {Brzezicki}}, \bibinfo {author} {\bibfnamefont {J.}~\bibnamefont
  {Dziarmaga}}, \ and\ \bibinfo {author} {\bibfnamefont {A.~M.}\ \bibnamefont
  {Ole\'s}},\ }\href@noop {} {\bibfield  {journal} {\bibinfo  {journal} {Phys.
  Rev. Lett.}\ }\textbf {\bibinfo {volume} {109}},\ \bibinfo {pages} {237201}
  (\bibinfo {year} {2012})}\BibitemShut {NoStop}%
\bibitem [{\citenamefont {Brzezicki}\ \emph {et~al.}(2013)\citenamefont
  {Brzezicki}, \citenamefont {Dziarmaga},\ and\ \citenamefont
  {Ole\'s}}]{Brz13}%
  \BibitemOpen
  \bibfield  {author} {\bibinfo {author} {\bibfnamefont {W.}~\bibnamefont
  {Brzezicki}}, \bibinfo {author} {\bibfnamefont {J.}~\bibnamefont
  {Dziarmaga}}, \ and\ \bibinfo {author} {\bibfnamefont {A.~M.}\ \bibnamefont
  {Ole\'s}},\ }\href@noop {} {\bibfield  {journal} {\bibinfo  {journal} {Phys.
  Rev. B}\ }\textbf {\bibinfo {volume} {87}},\ \bibinfo {pages} {064407}
  (\bibinfo {year} {2013})}\BibitemShut {NoStop}%
\bibitem [{\citenamefont {Feiner}\ and\ \citenamefont {Ole\'s}(1999)}]{Fei99}%
  \BibitemOpen
  \bibfield  {author} {\bibinfo {author} {\bibfnamefont {L.~F.}\ \bibnamefont
  {Feiner}}\ and\ \bibinfo {author} {\bibfnamefont {A.~M.}\ \bibnamefont
  {Ole\'s}},\ }\href@noop {} {\bibfield  {journal} {\bibinfo  {journal} {Phys.
  Rev. B}\ }\textbf {\bibinfo {volume} {59}},\ \bibinfo {pages} {3295}
  (\bibinfo {year} {1999})}\BibitemShut {NoStop}%
\bibitem [{\citenamefont {Feiner}\ and\ \citenamefont {Ole\'s}(2005)}]{Fei05}%
  \BibitemOpen
  \bibfield  {author} {\bibinfo {author} {\bibfnamefont {L.~F.}\ \bibnamefont
  {Feiner}}\ and\ \bibinfo {author} {\bibfnamefont {A.~M.}\ \bibnamefont
  {Ole\'s}},\ }\href@noop {} {\bibfield  {journal} {\bibinfo  {journal} {Phys.
  Rev. B}\ }\textbf {\bibinfo {volume} {71}},\ \bibinfo {pages} {144422}
  (\bibinfo {year} {2005})}\BibitemShut {NoStop}%
\bibitem [{\citenamefont {Khaliullin}\ \emph {et~al.}(2001)\citenamefont
  {Khaliullin}, \citenamefont {Horsch},\ and\ \citenamefont {Ole\'s}}]{Kha01}%
  \BibitemOpen
  \bibfield  {author} {\bibinfo {author} {\bibfnamefont {G.}~\bibnamefont
  {Khaliullin}}, \bibinfo {author} {\bibfnamefont {P.}~\bibnamefont {Horsch}},
  \ and\ \bibinfo {author} {\bibfnamefont {A.~M.}\ \bibnamefont {Ole\'s}},\
  }\href@noop {} {\bibfield  {journal} {\bibinfo  {journal} {Phys. Rev. Lett.}\
  }\textbf {\bibinfo {volume} {86}},\ \bibinfo {pages} {3879} (\bibinfo {year}
  {2001})}\BibitemShut {NoStop}%
\bibitem [{\citenamefont {Horsch}\ \emph {et~al.}(2008)\citenamefont {Horsch},
  \citenamefont {Ole\'s}, \citenamefont {Feiner},\ and\ \citenamefont
  {Khaliullin}}]{Hor08}%
  \BibitemOpen
  \bibfield  {author} {\bibinfo {author} {\bibfnamefont {P.}~\bibnamefont
  {Horsch}}, \bibinfo {author} {\bibfnamefont {A.~M.}\ \bibnamefont {Ole\'s}},
  \bibinfo {author} {\bibfnamefont {L.~F.}\ \bibnamefont {Feiner}}, \ and\
  \bibinfo {author} {\bibfnamefont {G.}~\bibnamefont {Khaliullin}},\
  }\href@noop {} {\bibfield  {journal} {\bibinfo  {journal} {Phys. Rev. Lett.}\
  }\textbf {\bibinfo {volume} {100}},\ \bibinfo {pages} {167205} (\bibinfo
  {year} {2008})}\BibitemShut {NoStop}%
\bibitem [{\citenamefont {Ole\'s}\ \emph {et~al.}(2006)\citenamefont {Ole\'s},
  \citenamefont {Horsch}, \citenamefont {Feiner},\ and\ \citenamefont
  {Khaliullin}}]{Ole06}%
  \BibitemOpen
  \bibfield  {author} {\bibinfo {author} {\bibfnamefont {A.~M.}\ \bibnamefont
  {Ole\'s}}, \bibinfo {author} {\bibfnamefont {P.}~\bibnamefont {Horsch}},
  \bibinfo {author} {\bibfnamefont {L.~F.}\ \bibnamefont {Feiner}}, \ and\
  \bibinfo {author} {\bibfnamefont {G.}~\bibnamefont {Khaliullin}},\
  }\href@noop {} {\bibfield  {journal} {\bibinfo  {journal} {Phys. Rev. Lett.}\
  }\textbf {\bibinfo {volume} {96}},\ \bibinfo {pages} {147205} (\bibinfo
  {year} {2006})}\BibitemShut {NoStop}%
\bibitem [{\citenamefont {You}(2012)}]{You12}%
  \BibitemOpen
  \bibfield  {author} {\bibinfo {author} {\bibfnamefont {W.-L.}\ \bibnamefont
  {You}}, {\bibfnamefont {A.~M.}\ \bibnamefont
  {Ole\'s}}, \ and\ \bibinfo {author} {\bibfnamefont {P.}~\bibnamefont {Horsch}},\ }\href@noop {} {\bibfield  {journal} {\bibinfo  {journal} {Phys.
  Rev. B}\ }\textbf {\bibinfo {volume} {86}},\ \bibinfo {pages} {094412}
  (\bibinfo {year} {2012})}\BibitemShut {NoStop}%
\bibitem [{\citenamefont {van~der Brink}\ \emph {et~al.}(1999)\citenamefont
  {van~der Brink}, \citenamefont {Horsch}, \citenamefont {Mack},\ and\
  \citenamefont {Ole\'s}}]{vdB99}%
  \BibitemOpen
  \bibfield  {author} {\bibinfo {author} {\bibfnamefont {J.}~\bibnamefont
  {van~der Brink}}, \bibinfo {author} {\bibfnamefont {P.}~\bibnamefont
  {Horsch}}, \bibinfo {author} {\bibfnamefont {F.}~\bibnamefont {Mack}}, \ and\
  \bibinfo {author} {\bibfnamefont {A.~M.}\ \bibnamefont {Ole\'s}},\
  }\href@noop {} {\bibfield  {journal} {\bibinfo  {journal} {Phys. Rev. B}\
  }\textbf {\bibinfo {volume} {59}},\ \bibinfo {pages} {6795} (\bibinfo {year}
  {1999})}\BibitemShut {NoStop}%
\bibitem [{\citenamefont {van~den Brink}(2004)}]{vdB04}%
  \BibitemOpen
  \bibfield  {author} {\bibinfo {author} {\bibfnamefont {J.}~\bibnamefont
  {van~den Brink}},\ }\href@noop {} {\bibfield  {journal} {\bibinfo  {journal}
  {New J. Phys.}\ }\textbf {\bibinfo {volume} {6}},\ \bibinfo {pages} {201}
  (\bibinfo {year} {2004})}\BibitemShut {NoStop}%
\bibitem [{\citenamefont {van Rynbach}\ \emph {et~al.}(2010)\citenamefont {van
  Rynbach}, \citenamefont {Todo},\ and\ \citenamefont {Trebst}}]{Ryn10}%
  \BibitemOpen
  \bibfield  {author} {\bibinfo {author} {\bibfnamefont {A.}~\bibnamefont {van
  Rynbach}}, \bibinfo {author} {\bibfnamefont {S.}~\bibnamefont {Todo}}, \ and\
  \bibinfo {author} {\bibfnamefont {S.}~\bibnamefont {Trebst}},\ }\href@noop {}
  {\bibfield  {journal} {\bibinfo  {journal} {Phys. Rev. Lett.}\ }\textbf
  {\bibinfo {volume} {105}},\ \bibinfo {pages} {146402} (\bibinfo {year}
  {2010})}\BibitemShut {NoStop}%
\bibitem [{\citenamefont {Trousselet}\ \emph
  {et~al.}(2012{\natexlab{a}})\citenamefont {Trousselet}, \citenamefont
  {Ralko},\ and\ \citenamefont {Ole\'s}}]{Ral12}%
  \BibitemOpen
  \bibfield  {author} {\bibinfo {author} {\bibfnamefont {F.}~\bibnamefont
  {Trousselet}}, \bibinfo {author} {\bibfnamefont {A.}~\bibnamefont {Ralko}}, \
  and\ \bibinfo {author} {\bibfnamefont {A.~M.}\ \bibnamefont {Ole\'s}},\
  }\href@noop {} {\bibfield  {journal} {\bibinfo  {journal} {Phys. Rev. B}\
  }\textbf {\bibinfo {volume} {86}},\ \bibinfo {pages} {014432} (\bibinfo
  {year} {2012}{\natexlab{a}})}\BibitemShut {NoStop}%
\bibitem [{\citenamefont {Nussinov}\ and\ \citenamefont {van~den
  Brink}(2013)}]{vdB13}%
  \BibitemOpen
  \bibfield  {author} {\bibinfo {author} {\bibfnamefont {Z.}~\bibnamefont
  {Nussinov}}\ and\ \bibinfo {author} {\bibfnamefont {J.}~\bibnamefont
  {van~den Brink}},\ }\href@noop {} {\bibfield  {journal} {\bibinfo  {journal}
  {arXiv:1303.5922 (unpublished)}\ } (\bibinfo
  {year} {2013})}\BibitemShut {NoStop}%
\bibitem [{\citenamefont {Kugel}\ and\ \citenamefont {Khomskii}(1982)}]{Kug82}%
  \BibitemOpen
  \bibfield  {author} {\bibinfo {author} {\bibfnamefont {K.~I.}\ \bibnamefont
  {Kugel}}\ and\ \bibinfo {author} {\bibfnamefont {D.~I.}\ \bibnamefont
  {Khomskii}},\ }\href@noop {} {\bibfield  {journal} {\bibinfo  {journal} {Sov.
  Phys. Usp.}\ }\textbf {\bibinfo {volume} {25}},\ \bibinfo {pages} {231}
  (\bibinfo {year} {1982})}\ {\bibfield  {journal} {\bibinfo  {journal} {[Usp.
  Fiz. Nauk}\ }\textbf {\bibinfo {volume} {136}},\ \bibinfo {pages} {621}
  (\bibinfo {year} {1982})}]\BibitemShut {NoStop}%
\bibitem [{\citenamefont {Nussinov}\ \emph {et~al.}(2004)\citenamefont
  {Nussinov}, \citenamefont {Biskup}, \citenamefont {Chayes},\ and\
  \citenamefont {van~den Brink}}]{Nus04}%
  \BibitemOpen
  \bibfield  {author} {\bibinfo {author} {\bibfnamefont {Z.}~\bibnamefont
  {Nussinov}}, \bibinfo {author} {\bibfnamefont {M.}~\bibnamefont {Biskup}},
  \bibinfo {author} {\bibfnamefont {L.}~\bibnamefont {Chayes}}, \ and\ \bibinfo
  {author} {\bibfnamefont {J.}~\bibnamefont {van~den Brink}},\ }\href@noop {}
  {\bibfield  {journal} {\bibinfo  {journal} {Europhys. Lett.}\ }\textbf
  {\bibinfo {volume} {67}},\ \bibinfo {pages} {990} (\bibinfo {year}
  {2004})}\BibitemShut {NoStop}%
\bibitem [{\citenamefont {Nussinov}\ and\ \citenamefont
  {Fradkin}(2005)}]{Nus05}%
  \BibitemOpen
  \bibfield  {author} {\bibinfo {author} {\bibfnamefont {Z.}~\bibnamefont
  {Nussinov}}\ and\ \bibinfo {author} {\bibfnamefont {E.}~\bibnamefont
  {Fradkin}},\ }\href@noop {} {\bibfield  {journal} {\bibinfo  {journal} {Phys.
  Rev. B}\ }\textbf {\bibinfo {volume} {71}},\ \bibinfo {pages} {195120}
  (\bibinfo {year} {2005})}\BibitemShut {NoStop}%
\bibitem [{\citenamefont {Xu}\ and\ \citenamefont {Moore}(2004)}]{Xu04}%
  \BibitemOpen
  \bibfield  {author} {\bibinfo {author} {\bibfnamefont {C.}~\bibnamefont
  {Xu}}\ and\ \bibinfo {author} {\bibfnamefont {J.~E.}\ \bibnamefont {Moore}},\
  }\href@noop {} {\bibfield  {journal} {\bibinfo  {journal} {Phys. Rev. Lett.}\
  }\textbf {\bibinfo {volume} {93}},\ \bibinfo {pages} {047003} (\bibinfo
  {year} {2004})}\BibitemShut {NoStop}%
\bibitem [{\citenamefont {Vidal}\ \emph {et~al.}(2009)\citenamefont {Vidal},
  \citenamefont {Thomale}, \citenamefont {Schmidt},\ and\ \citenamefont
  {Dusuel}}]{Vid09}%
  \BibitemOpen
  \bibfield  {author} {\bibinfo {author} {\bibfnamefont {J.}~\bibnamefont
  {Vidal}}, \bibinfo {author} {\bibfnamefont {R.}~\bibnamefont {Thomale}},
  \bibinfo {author} {\bibfnamefont {K.~P.}\ \bibnamefont {Schmidt}}, \ and\
  \bibinfo {author} {\bibfnamefont {S.}~\bibnamefont {Dusuel}},\ }\href@noop {}
  {\bibfield  {journal} {\bibinfo  {journal} {Phys. Rev. B}\ }\textbf {\bibinfo
  {volume} {80}},\ \bibinfo {pages} {081104} (\bibinfo {year}
  {2009})}\BibitemShut {NoStop}%
\bibitem [{\citenamefont {Cobanera}\ \emph {et~al.}(2010)\citenamefont
  {Cobanera}, \citenamefont {Ortiz},\ and\ \citenamefont {Nussinov}}]{Cob10}%
  \BibitemOpen
  \bibfield  {author} {\bibinfo {author} {\bibfnamefont {E.}~\bibnamefont
  {Cobanera}}, \bibinfo {author} {\bibfnamefont {G.}~\bibnamefont {Ortiz}}, \
  and\ \bibinfo {author} {\bibfnamefont {Z.}~\bibnamefont {Nussinov}},\
  }\href@noop {} {\bibfield  {journal} {\bibinfo  {journal} {Phys. Rev. Lett.}\
  }\textbf {\bibinfo {volume} {104}},\ \bibinfo {pages} {020402} (\bibinfo
  {year} {2010})}\BibitemShut {NoStop}%
\bibitem [{\citenamefont {Dou\c{c}ot}\ \emph {et~al.}(2005)\citenamefont
  {Dou\c{c}ot}, \citenamefont {Feigel'man}, \citenamefont {Ioffe},\ and\
  \citenamefont {Ioselevich}}]{Dou05}%
  \BibitemOpen
  \bibfield  {author} {\bibinfo {author} {\bibfnamefont {B.}~\bibnamefont
  {Dou\c{c}ot}}, \bibinfo {author} {\bibfnamefont {M.~V.}\ \bibnamefont
  {Feigel'man}}, \bibinfo {author} {\bibfnamefont {L.~B.}\ \bibnamefont
  {Ioffe}}, \ and\ \bibinfo {author} {\bibfnamefont {A.~S.}\ \bibnamefont
  {Ioselevich}},\ }\href@noop {} {\bibfield  {journal} {\bibinfo  {journal}
  {Phys. Rev. B}\ }\textbf {\bibinfo {volume} {71}},\ \bibinfo {pages} {024505}
  (\bibinfo {year} {2005})}\BibitemShut {NoStop}%
\bibitem [{\citenamefont {Gladchenko}\ \emph {et~al.}(2009)\citenamefont
  {Gladchenko}, \citenamefont {Olaya}, \citenamefont {Dupont-Ferrier},
  \citenamefont {Dou\c{c}ot}, \citenamefont {Ioffe},\ and\ \citenamefont
  {Gershenson}}]{Gla09}%
  \BibitemOpen
  \bibfield  {author} {\bibinfo {author} {\bibfnamefont {S.}~\bibnamefont
  {Gladchenko}}, \bibinfo {author} {\bibfnamefont {D.}~\bibnamefont {Olaya}},
  \bibinfo {author} {\bibfnamefont {E.}~\bibnamefont {Dupont-Ferrier}},
  \bibinfo {author} {\bibfnamefont {B.}~\bibnamefont {Dou\c{c}ot}}, \bibinfo
  {author} {\bibfnamefont {L.~B.}\ \bibnamefont {Ioffe}}, \ and\ \bibinfo
  {author} {\bibfnamefont {M.~E.}\ \bibnamefont {Gershenson}},\ }\href@noop {}
  {\bibfield  {journal} {\bibinfo  {journal} {J. Phys. Soc. Jpn.}\ }\textbf
  {\bibinfo {volume} {96}},\ \bibinfo {pages} {1606} (\bibinfo {year}
  {2009})}\BibitemShut {NoStop}%
\bibitem [{\citenamefont {Milman}\ \emph {et~al.}(2007)\citenamefont {Milman},
  \citenamefont {Maineult}, \citenamefont {Guibal}, \citenamefont {Guidoni},
  \citenamefont {Dou\c{c}ot}, \citenamefont {Ioffe},\ and\ \citenamefont
  {Coudreau}}]{Mil07}%
  \BibitemOpen
  \bibfield  {author} {\bibinfo {author} {\bibfnamefont {P.}~\bibnamefont
  {Milman}}, \bibinfo {author} {\bibfnamefont {W.}~\bibnamefont {Maineult}},
  \bibinfo {author} {\bibfnamefont {S.}~\bibnamefont {Guibal}}, \bibinfo
  {author} {\bibfnamefont {L.}~\bibnamefont {Guidoni}}, \bibinfo {author}
  {\bibfnamefont {B.}~\bibnamefont {Dou\c{c}ot}}, \bibinfo {author}
  {\bibfnamefont {L.}~\bibnamefont {Ioffe}}, \ and\ \bibinfo {author}
  {\bibfnamefont {T.}~\bibnamefont {Coudreau}},\ }\href@noop {} {\bibfield
  {journal} {\bibinfo  {journal} {Phys. Rev. Lett.}\ }\textbf {\bibinfo
  {volume} {99}},\ \bibinfo {pages} {020503} (\bibinfo {year}
  {2007})}\BibitemShut {NoStop}%
\bibitem [{\citenamefont {Khomskii}\ and\ \citenamefont
  {Mostovoy}(2003)}]{Kho03}%
  \BibitemOpen
  \bibfield  {author} {\bibinfo {author} {\bibfnamefont {D.~I.}\ \bibnamefont
  {Khomskii}}\ and\ \bibinfo {author} {\bibfnamefont {M.~V.}\ \bibnamefont
  {Mostovoy}},\ }\href@noop {} {\bibfield  {journal} {\bibinfo  {journal} {J.
  Phys. A: Math. Gen.}\ }\textbf {\bibinfo {volume} {36}},\ \bibinfo {pages}
  {9197} (\bibinfo {year} {2003})}\BibitemShut {NoStop}%
\bibitem [{\citenamefont {Dorier}\ \emph {et~al.}(2005)\citenamefont {Dorier},
  \citenamefont {Becca},\ and\ \citenamefont {Mila}}]{Mil05}%
  \BibitemOpen
  \bibfield  {author} {\bibinfo {author} {\bibfnamefont {J.}~\bibnamefont
  {Dorier}}, \bibinfo {author} {\bibfnamefont {F.}~\bibnamefont {Becca}}, \
  and\ \bibinfo {author} {\bibfnamefont {F.}~\bibnamefont {Mila}},\ }\href@noop
  {} {\bibfield  {journal} {\bibinfo  {journal} {Phys. Rev. B}\ }\textbf
  {\bibinfo {volume} {72}},\ \bibinfo {pages} {024448} (\bibinfo {year}
  {2005})}\BibitemShut {NoStop}%
\bibitem [{\citenamefont {You}\ \emph {et~al.}(2010)\citenamefont {You},
  \citenamefont {Tian},\ and\ \citenamefont {Lin}}]{You10}%
  \BibitemOpen
  \bibfield  {author} {\bibinfo {author} {\bibfnamefont {W.-L.}\ \bibnamefont
  {You}}, \bibinfo {author} {\bibfnamefont {G.-S.}\ \bibnamefont {Tian}}, \
  and\ \bibinfo {author} {\bibfnamefont {H.-Q.}\ \bibnamefont {Lin}},\
  }\href@noop {} {\bibfield  {journal} {\bibinfo  {journal} {J. Phys. A: Math. Gen.}\
  }\textbf {\bibinfo {volume} {43}},\ \bibinfo {pages} {275001} (\bibinfo
  {year} {2010})}\BibitemShut {NoStop}%
\bibitem [{\citenamefont {Chen}\ \emph {et~al.}(2007)\citenamefont {Chen},
  \citenamefont {Fang}, \citenamefont {Hu},\ and\ \citenamefont {Yao}}]{Che07}%
  \BibitemOpen
  \bibfield  {author} {\bibinfo {author} {\bibfnamefont {H.-D.}\ \bibnamefont
  {Chen}}, \bibinfo {author} {\bibfnamefont {C.}~\bibnamefont {Fang}}, \bibinfo
  {author} {\bibfnamefont {J.}~\bibnamefont {Hu}}, \ and\ \bibinfo {author}
  {\bibfnamefont {H.}~\bibnamefont {Yao}},\ }\href@noop {} {\bibfield
  {journal} {\bibinfo  {journal} {Phys. Rev. B}\ }\textbf {\bibinfo {volume}
  {75}},\ \bibinfo {pages} {144401} (\bibinfo {year} {2007})}\BibitemShut
  {NoStop}%
\bibitem [{\citenamefont {Or\'us}\ \emph {et~al.}(2009)\citenamefont {Or\'us},
  \citenamefont {Doherty},\ and\ \citenamefont {Vidal}}]{Oru09}%
  \BibitemOpen
  \bibfield  {author} {\bibinfo {author} {\bibfnamefont {R.}~\bibnamefont
  {Or\'us}}, \bibinfo {author} {\bibfnamefont {A.~C.}\ \bibnamefont {Doherty}},
  \ and\ \bibinfo {author} {\bibfnamefont {G.}~\bibnamefont {Vidal}},\
  }\href@noop {} {\bibfield  {journal} {\bibinfo  {journal} {Phys. Rev. Lett.}\
  }\textbf {\bibinfo {volume} {102}},\ \bibinfo {pages} {077203} (\bibinfo
  {year} {2009})}\BibitemShut {NoStop}%
\bibitem [{\citenamefont {Wenzel}\ and\ \citenamefont {Janke}(2008)}]{Wen08}%
  \BibitemOpen
  \bibfield  {author} {\bibinfo {author} {\bibfnamefont {S.}~\bibnamefont
  {Wenzel}}\ and\ \bibinfo {author} {\bibfnamefont {W.}~\bibnamefont {Janke}},\
  }\href@noop {} {\bibfield  {journal} {\bibinfo  {journal} {Phys. Rev. B}\
  }\textbf {\bibinfo {volume} {78}},\ \bibinfo {pages} {064402} (\bibinfo
  {year} {2008})}\BibitemShut {NoStop}%
\bibitem [{\citenamefont {Trousselet}\ \emph {et~al.}(2010)\citenamefont
  {Trousselet}, \citenamefont {Ole\'s},\ and\ \citenamefont {Horsch}}]{Tro10}%
  \BibitemOpen
  \bibfield  {author} {\bibinfo {author} {\bibfnamefont {F.}~\bibnamefont
  {Trousselet}}, \bibinfo {author} {\bibfnamefont {A.~M.}\ \bibnamefont
  {Ole\'s}}, \ and\ \bibinfo {author} {\bibfnamefont {P.}~\bibnamefont
  {Horsch}},\ }\href@noop {} {\bibfield  {journal} {\bibinfo  {journal}
  {Europhys. Lett.}\ }\textbf {\bibinfo {volume} {91}},\ \bibinfo {pages}
  {40005} (\bibinfo {year} {2010})}\BibitemShut {NoStop}%
\bibitem [{\citenamefont {Trousselet}\ \emph
  {et~al.}(2012{\natexlab{b}})\citenamefont {Trousselet}, \citenamefont
  {Ole\'s},\ and\ \citenamefont {Horsch}}]{Tro12}%
  \BibitemOpen
  \bibfield  {author} {\bibinfo {author} {\bibfnamefont {F.}~\bibnamefont
  {Trousselet}}, \bibinfo {author} {\bibfnamefont {A.~M.}\ \bibnamefont
  {Ole\'s}}, \ and\ \bibinfo {author} {\bibfnamefont {P.}~\bibnamefont
  {Horsch}},\ }\href@noop {} {\bibfield  {journal} {\bibinfo  {journal} {Phys.
  Rev. B}\ }\textbf {\bibinfo {volume} {86}},\ \bibinfo {pages} {134412}
  (\bibinfo {year} {2012}{\natexlab{b}})}\BibitemShut {NoStop}%
\bibitem [{\citenamefont {Cincio}\ \emph {et~al.}(2010)\citenamefont {Cincio},
  \citenamefont {Dziarmaga},\ and\ \citenamefont {Ole\'s}}]{Cin10}%
  \BibitemOpen
  \bibfield  {author} {\bibinfo {author} {\bibfnamefont {L.}~\bibnamefont
  {Cincio}}, \bibinfo {author} {\bibfnamefont {J.}~\bibnamefont {Dziarmaga}}, \
  and\ \bibinfo {author} {\bibfnamefont {A.~M.}\ \bibnamefont {Ole\'s}},\
  }\href@noop {} {\bibfield  {journal} {\bibinfo  {journal} {Phys. Rev. B}\
  }\textbf {\bibinfo {volume} {82}},\ \bibinfo {pages} {104416} (\bibinfo
  {year} {2010})}\BibitemShut {NoStop}%
\bibitem [{\citenamefont {Brzezicki}\ \emph {et~al.}(2007)\citenamefont
  {Brzezicki}, \citenamefont {Dziarmaga},\ and\ \citenamefont {Ole\'s}}]{wb07}%
  \BibitemOpen
  \bibfield  {author} {\bibinfo {author} {\bibfnamefont {W.}~\bibnamefont
  {Brzezicki}}, \bibinfo {author} {\bibfnamefont {J.}~\bibnamefont
  {Dziarmaga}}, \ and\ \bibinfo {author} {\bibfnamefont {A.~M.}\ \bibnamefont
  {Ole\'s}},\ }\href@noop {} {\bibfield  {journal} {\bibinfo  {journal} {Phys.
  Rev. B}\ }\textbf {\bibinfo {volume} {75}},\ \bibinfo {pages} {134415}
  (\bibinfo {year} {2007})}\BibitemShut {NoStop}%
\bibitem [{\citenamefont {Brzezicki}\ and\ \citenamefont
  {Ole\'s}(2009{\natexlab{a}})}]{wb09_act}%
  \BibitemOpen
  \bibfield  {author} {\bibinfo {author} {\bibfnamefont {W.}~\bibnamefont
  {Brzezicki}}\ and\ \bibinfo {author} {\bibfnamefont {A.~M.}\ \bibnamefont
  {Ole\'s}},\ }\href@noop {} {\bibfield  {journal} {\bibinfo  {journal} {Acta
  Phys. Pol. A}\ }\textbf {\bibinfo {volume} {115}},\ \bibinfo {pages} {162}
  (\bibinfo {year} {2009}{\natexlab{a}})}\BibitemShut {NoStop}%
\bibitem [{\citenamefont {Perk}\ \emph {et~al.}(1975)\citenamefont {Perk},
  \citenamefont {Capel}, \citenamefont {Zuilhof},\ and\ \citenamefont
  {Siskens}}]{Per75}%
  \BibitemOpen
  \bibfield  {author} {\bibinfo {author} {\bibfnamefont {J.~H.~H.}\
  \bibnamefont {Perk}}, \bibinfo {author} {\bibfnamefont {H.~W.}\ \bibnamefont
  {Capel}}, \bibinfo {author} {\bibfnamefont {M.~J.}\ \bibnamefont {Zuilhof}},
  \ and\ \bibinfo {author} {\bibfnamefont {T.~J.}\ \bibnamefont {Siskens}},\
  }\href@noop {} {\bibfield  {journal} {\bibinfo  {journal} {Physica A}\
  }\textbf {\bibinfo {volume} {81}},\ \bibinfo {pages} {319} (\bibinfo {year}
  {1975})}\BibitemShut {NoStop}%
\bibitem [{\citenamefont {Eriksson}\ and\ \citenamefont
  {Johannesson}(2009)}]{Eri09}%
  \BibitemOpen
  \bibfield  {author} {\bibinfo {author} {\bibfnamefont {E.}~\bibnamefont
  {Eriksson}}\ and\ \bibinfo {author} {\bibfnamefont {H.}~\bibnamefont
  {Johannesson}},\ }\href@noop {} {\bibfield  {journal} {\bibinfo  {journal}
  {Phys. Rev. B}\ }\textbf {\bibinfo {volume} {79}},\ \bibinfo {pages} {224424}
  (\bibinfo {year} {2009})}\BibitemShut {NoStop}%
\bibitem [{\citenamefont {You}\ and\ \citenamefont {Tian}(2008)}]{You08}%
  \BibitemOpen
  \bibfield  {author} {\bibinfo {author} {\bibfnamefont {W.-L.}\ \bibnamefont
  {You}}\ and\ \bibinfo {author} {\bibfnamefont {G.-S.}\ \bibnamefont {Tian}},\
  }\href@noop {} {\bibfield  {journal} {\bibinfo  {journal} {Phys. Rev. B}\
  }\textbf {\bibinfo {volume} {78}},\ \bibinfo {pages} {184406} (\bibinfo
  {year} {2008})}\BibitemShut {NoStop}%
\bibitem [{\citenamefont {Brzezicki}\ and\ \citenamefont
  {Ole\'s}(2010{\natexlab{a}})}]{wb10}%
  \BibitemOpen
  \bibfield  {author} {\bibinfo {author} {\bibfnamefont {W.}~\bibnamefont
  {Brzezicki}}\ and\ \bibinfo {author} {\bibfnamefont {A.~M.}\ \bibnamefont
  {Ole\'s}},\ }\href@noop {} {\bibfield  {journal} {\bibinfo  {journal} {Phys.
  Rev. B}\ }\textbf {\bibinfo {volume} {82}},\ \bibinfo {pages} {060401}
  (\bibinfo {year} {2010}{\natexlab{a}})}\BibitemShut {NoStop}%
\bibitem [{\citenamefont {Brzezicki}\ and\ \citenamefont
  {Ole\'s}(2010{\natexlab{b}})}]{wb10_icm}%
  \BibitemOpen
  \bibfield  {author} {\bibinfo {author} {\bibfnamefont {W.}~\bibnamefont
  {Brzezicki}}\ and\ \bibinfo {author} {\bibfnamefont {A.~M.}\ \bibnamefont
  {Ole\'s}},\ }\href@noop {} {\bibfield  {journal} {\bibinfo  {journal} {J.
  Phys.: Conf. Ser.}\ }\textbf {\bibinfo {volume} {200}},\ \bibinfo {pages}
  {012017} (\bibinfo {year} {2010}{\natexlab{b}})}\BibitemShut {NoStop}%
\bibitem [{\citenamefont {Brzezicki}(2011)}]{wb11_vi3}%
  \BibitemOpen
  \bibfield  {author} {\bibinfo {author} {\bibfnamefont {W.}~\bibnamefont
  {Brzezicki}},\ }\href@noop {} {\emph {\bibinfo {title} {Lectures on the
  Physics of Strongly Correlated Systems XV}}},\ \bibinfo {series} {AIP
  Conference Proceedings}, Vol.\ \bibinfo {volume} \textbf{1419}\ (\bibinfo
  {publisher} {AIP},\ \bibinfo {address} {New York},\ \bibinfo {year} {2011})\
  \bibinfo {note} {pp. 261-265}\BibitemShut {NoStop}%
\bibitem [{\citenamefont {Nakano}\ and\ \citenamefont {Sakai}(2011)}]{Nak11}%
  \BibitemOpen
  \bibfield  {author} {\bibinfo {author} {\bibfnamefont {H.}~\bibnamefont
  {Nakano}}\ and\ \bibinfo {author} {\bibfnamefont {T.}~\bibnamefont {Sakai}},\
  }\href@noop {} {\bibfield  {journal} {\bibinfo  {journal} {J. Phys. Soc.
  Jpn.}\ }\textbf {\bibinfo {volume} {80}},\ \bibinfo {pages} {053704}
  (\bibinfo {year} {2011})}\BibitemShut {NoStop}%
\bibitem [{\citenamefont {Brzezicki}\ and\ \citenamefont
  {Ole\'s}(2009{\natexlab{b}})}]{wb09}%
  \BibitemOpen
  \bibfield  {author} {\bibinfo {author} {\bibfnamefont {W.}~\bibnamefont
  {Brzezicki}}\ and\ \bibinfo {author} {\bibfnamefont {A.~M.}\ \bibnamefont
  {Ole\'s}},\ }\href@noop {} {\bibfield  {journal} {\bibinfo  {journal} {Phys.
  Rev. B}\ }\textbf {\bibinfo {volume} {80}},\ \bibinfo {pages} {014405}
  (\bibinfo {year} {2009}{\natexlab{b}})}\BibitemShut {NoStop}%
\bibitem [{\citenamefont {Brzezicki}(2010)}]{wb10_vi2}%
  \BibitemOpen
  \bibfield  {author} {\bibinfo {author} {\bibfnamefont {W.}~\bibnamefont
  {Brzezicki}},\ }\href@noop {} {\emph {\bibinfo {title} {Lectures on the
  Physics of Strongly Correlated Systems XIV}}},\ \bibinfo {series} {AIP
  Conference Proceedings}, Vol.\ \bibinfo {volume} {\textbf{1297}}\ (\bibinfo
  {publisher} {AIP},\ \bibinfo {address} {New York},\ \bibinfo {year} {2010})\
  \bibinfo {note} {pp. 407-411}\BibitemShut {NoStop}%
\bibitem [{\citenamefont {Weisse}\ and\ \citenamefont {Fehske}(2008)}]{Feh08}%
  \BibitemOpen
  \bibfield  {author} {\bibinfo {author} {\bibfnamefont {A.}~\bibnamefont
  {Weisse}}\ and\ \bibinfo {author} {\bibfnamefont {H.}~\bibnamefont
  {Fehske}},\ }\href@noop {} {\emph {\bibinfo {title} {Computational Many
  Particle Physics}}},\ \bibinfo {series} {Lect. Notes Phys.}, Vol.\ \bibinfo
  {volume} {\textbf{739}}\ (\bibinfo  {publisher} {Springer},\ \bibinfo {address}
  {Berlin},\ \bibinfo {year} {2008})\ \bibinfo {note} {pp. 545-577}\BibitemShut
  {NoStop}%
\bibitem [{\citenamefont {Brzezicki}\ and\ \citenamefont
  {Ole\'s}(2011)}]{wb11}%
  \BibitemOpen
  \bibfield  {author} {\bibinfo {author} {\bibfnamefont {W.}~\bibnamefont
  {Brzezicki}}\ and\ \bibinfo {author} {\bibfnamefont {A.~M.}\ \bibnamefont
  {Ole\'s}},\ }\href@noop {} {\bibfield  {journal} {\bibinfo  {journal} {Phys.
  Rev. B}\ }\textbf {\bibinfo {volume} {83}},\ \bibinfo {pages} {214408}
  (\bibinfo {year} {2011})}\BibitemShut {NoStop}%
\bibitem [{\citenamefont {Park}\ \emph {et~al.}(2010)\citenamefont {Park},
  \citenamefont {Lee},\ and\ \citenamefont {Lee}}]{Par10}%
  \BibitemOpen
  \bibfield  {author} {\bibinfo {author} {\bibfnamefont {T.~Y.}\ \bibnamefont
  {Park}}, \bibinfo {author} {\bibfnamefont {Y.~C.}\ \bibnamefont {Lee}}, \
  and\ \bibinfo {author} {\bibfnamefont {J.-W.}\ \bibnamefont {Lee}},\
  }\href@noop {} {\bibfield  {journal} {\bibinfo  {journal} {J. Korean Phys.
  Soc.}\ }\textbf {\bibinfo {volume} {56}},\ \bibinfo {pages} {1011} (\bibinfo
  {year} {2010})}\BibitemShut {NoStop}%
\end{thebibliography}

\end{document}